\documentclass[letterpaper,journal]{IEEEtran}
\usepackage{graphicx}
\usepackage{amsmath,amsfonts,amssymb}
\usepackage{cite}
\usepackage{algorithmic}
\usepackage{algorithm}
\usepackage{array}
\usepackage{textcomp}
\usepackage{stfloats}
\usepackage{url}
\usepackage{verbatim}
\usepackage{makecell}
\usepackage{multirow}
\usepackage{xcolor}
\usepackage{colortbl}
\usepackage{fontawesome}
\usepackage{threeparttable}
\usepackage{nomencl}
\usepackage{caption}
\usepackage[caption=false,font=normalsize,labelfont=sf,textfont=sf]{subfig}
\makenomenclature
\begin{document}
\title{Stacked Intelligent Metasurface-Aided Wave-Domain Signal Processing: From Communications to Sensing and Computing}
\author{Jiancheng An,~\IEEEmembership{Senior Member,~IEEE}, Chau Yuen,~\IEEEmembership{Fellow,~IEEE}, Marco Di Renzo,~\IEEEmembership{Fellow,~IEEE},\\Mehdi Bennis,~\IEEEmembership{Fellow,~IEEE}, and M\'erouane Debbah,~\IEEEmembership{Fellow,~IEEE}, and Lajos Hanzo,~\IEEEmembership{Life Fellow,~IEEE}
\thanks{This work was supported by the MOE (Ministry of Education, Singapore), under MOE Tier 2 Award number T2EP50124-0032. The work of M. Di Renzo was supported in part by the European Research Council (ERC) under the European Union’s Horizon Europe Programme WePhICom (grant agreement number 101225119), as well as by the European Union through the Horizon Europe project COVER under grant agreement number 101086228, the Horizon Europe project UNITE under grant agreement number 101129618, the Horizon Europe project INSTINCT under grant agreement number 101139161, and the Horizon Europe project TWIN6G under grant agreement number 101182794, as well as by the Agence Nationale de la Recherche (ANR) through the France 2030 project ANR-PEPR Networks of the Future under grant agreements NF-PERSEUS 22-PEFT-004, NF-YACARI 22-PEFT-0005, NF-SYSTERA 22-PEFT-0006, NF-Founds 22-PEFT-0010, ANR-24-PEFT-0001, and by the CHIST-ERA project PASSIONATE under grant agreements CHIST-ERA-22-WAI-04 and ANR-23-CHR4-0003-01. Also, the work of M. Di Renzo was supported in part by the Engineering and Physical Sciences Research Council (EPSRC), part of UK Research and Innovation, and the UK Department of Science, Innovation and Technology through the CHEDDAR Telecom Hub under grant EP/Y037421/1, through the HASC Telecom Hub under grant EP/Y037197/1, and through the TITAN Telecom Hub under grant EP/Y037243/1. \emph{(Corresponding Author: Chau Yuen)}}
\thanks{J. An and C. Yuen are with the School of Electrical and Electronics Engineering, Nanyang Technological University (NTU), Singapore 639798 (e-mail: jiancheng\_an@163.com, chau.yuen@ntu.edu.sg).}
\thanks{M. Di Renzo is with CNRS and CentraleSup\'elec, Institute of Electronics and Digital Technologies (IETR), Avenue de la Boulaie, 35576 Cesson-S\'evign\'e, France (marco.direnzo@centralesupelec.fr), and with King's College London, Department of Engineering - Centre for Telecommunications Research, WC2R 2LS London, United Kingdom (marco.di\_renzo@kcl.ac.uk).}
\thanks{M. Bennis is with the Centre for Wireless Communications, University of Oulu, 90570 Oulu, Finland (e-mail: mehdi.bennis@oulu.fi).}
\thanks{M. Debbah is with the Research Institute for Digital Future, Khalifa University, 127788 Abu Dhabi, UAE (email: merouane.debbah@ku.ac.ae) and also with CentraleSupelec, University Paris-Saclay, 91192 Gif-sur-Yvette, France.}
\thanks{L. Hanzo is with the School of Electronics and Computer Science, University of Southampton, Southampton SO17 1BJ, U.K. (e-mail: lh@ecs.soton.ac.uk).}\vspace{-1cm}}
\markboth{DRAFT}{DRAFT}
\maketitle
\begin{abstract}
Artificial neural networks possess remarkable capabilities for abstract feature extraction, while electromagnetic computing leverages wave propagation to execute complex mathematical operations. Concurrently, metasurfaces engineered from subwavelength meta-atoms offer unprecedented control over electromagnetic wavefronts. Synthesizing these three cutting-edge fields has sparked significant interest in developing electromagnetic neural networks via stacked intelligent metasurface (SIM) technology, which aims to execute diverse signal processing tasks directly within the wave domain. By enabling direct processing of information-carrying electromagnetic waves, SIMs offer a promising paradigm for high-speed, massively parallel, and low-power signal processing. This article provides a comprehensive overview of SIM technology, beginning with its evolutionary trajectory. We then delve into its theoretical foundations and examine state-of-the-art SIM hardware prototypes. Furthermore, we analyze the optimization and training strategies devised to configure SIM functionalities from two distinct perspectives. Additionally, the diverse applications of SIM technology across the communication, sensing, and computing domains are explored, supported by experimental evidence that highlights its ability to sustain multiple functions within a single device. Finally, we outline critical technical challenges to deploying SIMs in next-generation wireless networks and chart promising research directions to fully unlock their transformative potential.
\end{abstract}

\begin{IEEEkeywords}
Stacked intelligent metasurfaces (SIM), electromagnetic signal processing, electromagnetic neural networks, advanced transceiver design.
\end{IEEEkeywords}

\section{Introduction}
\subsection{Background and Motivation}
\IEEEPARstart{T}{he} past decade has witnessed tremendous advances in artificial intelligence (AI)~\cite{Nature_2015_LeCun_Deep}. In particular, multi-layer artificial neural networks (ANNs), inspired by the architecture of biological neural networks in the human brain, have demonstrated exceptional capabilities in a wide range of applications, such as computer vision~\cite{CVPR_2016_Szegedy_Rethinking}, speech recognition~\cite{SPM_2012_Hinton_Deep}, and natural language processing~\cite{TNNLS_2021_Otter_A}. An ANN is composed of artificial neurons that are typically aggregated into layers, with connections to other neurons in adjacent layers through synaptic edges. Each artificial neuron processes the signals it receives from connected neurons and sends a signal to the downstream neurons~\cite{Proc_2017_Sze_Efficient}. By training the connection weights of each edge in a data-driven manner, ANNs acquire the ability to learn abstract features of the input data, enabling them to handle complex tasks.

Analog electromagnetic computing (or wave-domain computing) has also gained considerable attention in recent years due to its ability to efficiently tackle complex mathematical operations, such as spatial integration, differentiation, and convolution, as well as solving equations by leveraging the propagation of waves within custom-designed material slabs or cavities~\cite{Sci_2014_Silva_Performing}. Electromagnetic computing has been found to significantly enhance the information processing efficiency compared to conventional computing paradigms, thanks to its unique characteristics of light-speed operation, low energy consumption, and scalability for massively parallel processing~\cite{NC_2025_Tzarouchis_Programmable}. Recent research studies have showcased the substantial potential of electromagnetic computing in various practical scenarios~\cite{Sci_2014_Silva_Performing}, particularly in high-volume real-time data processing scenarios, such as augmented reality and autonomous driving.

Additionally, the rapidly advancing field of engineered metamaterials has led to the emergence of programmable metasurfaces and information metasurfaces~\cite{LSA_2014_Cui_Coding, Proc_2022_Cheng_Reconfigurable}. A metasurface is essentially an artificial planar surface composed of periodically arranged subwavelength scattering meta-atoms~\cite{TWC_2025_An_Flexible, Proc_2022_Renzo_Communication}. By tuning external stimuli, such as mechanical deformation or bias voltage using field programmable gate arrays (FPGA), each individual meta-atom can independently modulate the phase and/or amplitude of incident electromagnetic waves~\cite{JSAC_2020_Renzo_Smart, Proc_2022_Alexandropoulos_Pervasive}. This breakthrough technology has opened up new possibilities for manipulating electromagnetic wavefronts at an unprecedented scale to achieve a wide range of functionalities, spanning from planar hologram generation to the customization of communication environments~\cite{arXiv_2024_Jiancheng_Emerging}.

\begin{figure*}[!t]
\centering
\includegraphics[width=18cm]{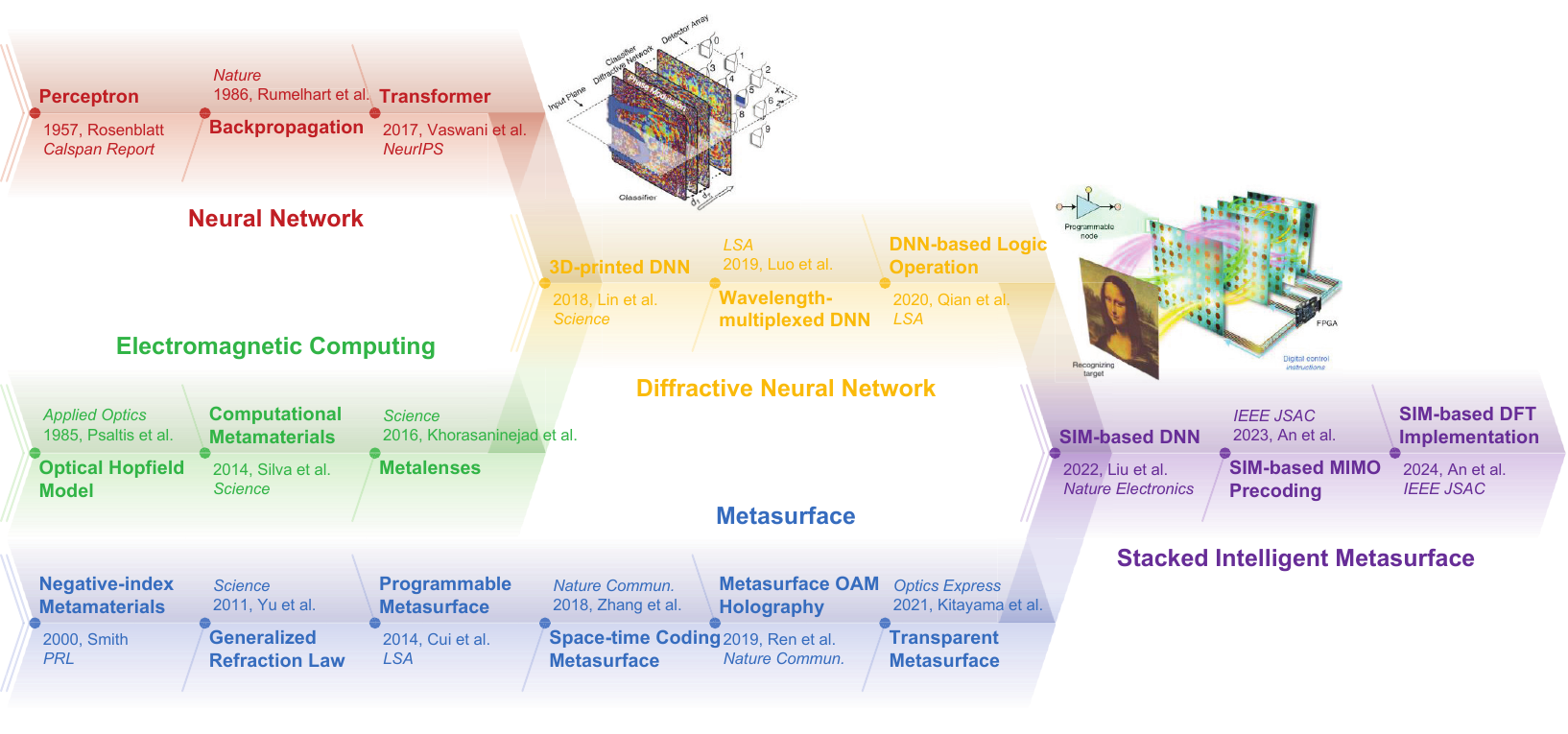}
\caption{The evolution timeline of the SIM technology.}
\label{fig_1}\vspace{-0.4cm}
\end{figure*}

The fusion of neural network architectures with electromagnetic computing has catalyzed the development of diffractive neural networks (DNNs)~\cite{Sci_2018_Lin_All}. As illustrated in Fig.~\ref{fig_1}, a DNN is implemented using a number of densely packed diffractive layers, with free-space propagation between them. In a dielectric diffractive layer, the thickness of each individual pixel (or neuron) is a learnable variable that can be adjusted to control the transmission coefficient~\cite{Sci_2018_Lin_All}. After adequate training and fabrication using three-dimensional (3D) printing techniques, a DNN can carry out specific functions through the interactions of electromagnetic waves within the diffractive layers. Various advanced DNN architectures have been reported in the recent literature, demonstrating impressive capabilities in many emerging applications, including image classification and reconstruction~\cite{SA_2021_Li_Spectrally}, real-time object detection~\cite{LSA_2022_Luo_Metasurface}, and matrix inversion~\cite{APR_2024_Chen_Diffractive}.

However, DNNs constructed with passive diffractive layers have a significant limitation: once they are fabricated, they cannot be retrained to perform other tasks. This restriction severely hinders their potential for widespread deployment. Fortunately, the emergence of programmable metasurfaces provides the missing piece of the puzzle to create reconfigurable DNNs~\cite{NE_2022_Liu_A}. As shown in Fig.~\ref{fig_1}, by stacking several intelligent metasurfaces into a single device, a new technology is conceived, which is referred to as a stacked intelligent metasurface (SIM)~\cite{JSAC_2022_An_Stacked, WC_2024_An_Stacked}. In a SIM, each intelligent metasurface functions as a layer of a DNN, and each programmable meta-atom is analogous to a neuron with learnable phase and amplitude responses, which can be adjusted to accommodate the requirements of diverse tasks in dynamic environments. Consequently, a SIM inherits the powerful representation capabilities of ANNs, the ultra-fast speed of electromagnetic computing, as well as the energy-efficient electromagnetic tuning capability of programmable metasurfaces~\cite{WC_2024_An_Stacked}.

\subsection{Organization}
SIMs represent a revolutionary and transformative technology that has the potential to fundamentally reshape the future of communication, sensing, and computing. In this article, we provide a comprehensive overview of SIM technology, examining its vast potential and the fundamental challenges in its practical applications. Specifically, our contributions are summarized below:
\begin{enumerate}
\item We commence by elucidating the fundamental SIM principles, encompassing its physical architecture and neural network counterparts. We further present a comprehensive review of advances in SIM prototype development. The configuration (or training, in DNN jargon) strategies for SIM are also examined from two different perspectives.
\item We critically appraise the applications of SIM in communication, sensing, and computing systems, presenting experimental findings that demonstrate the distinctive advantages of SIM technology. Additionally, the promising potential for integrating communication, sensing, and computing functionalities within a unified SIM framework is discussed to enable emerging applications.
\item We shed light on both the substantial opportunities and critical challenges inherent in translating the innovative SIM concept into a practical technology, addressing considerations in modeling, protocol design, and potential energy efficiency concerns. Furthermore, we provide insights into pathways for enhancing both the signal processing and inference capabilities of SIMs.
\end{enumerate}
\begin{figure}[!t]
\centering
\includegraphics[width=8.5cm]{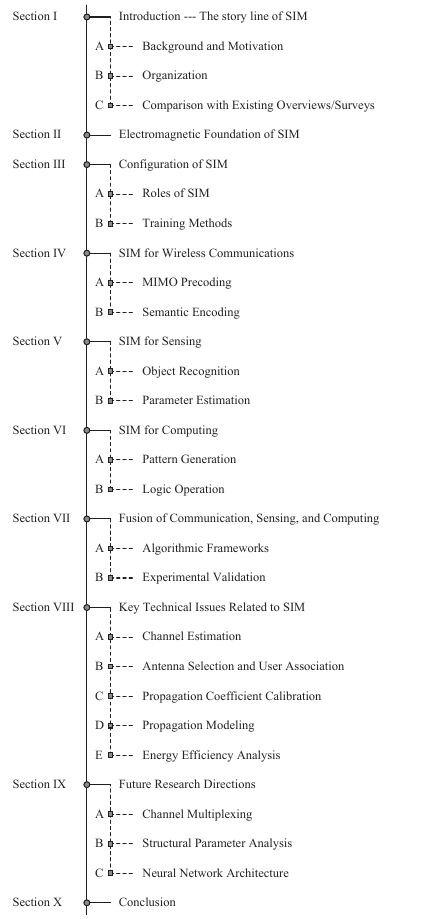}
\caption{The organization of this paper.}
\label{fig_2}\vspace{-0.4cm}
\end{figure}
To outline our exploration of SIM technology, the structure of this article is illustrated in Fig.~\ref{fig_2}.

\subsection{Comparison with Existing Overview/Survey}
As SIM has garnered growing research interest, the volume of related publications has expanded rapidly. To situate the present survey within this broader landscape, we provide a comparative discussion of the existing overviews and surveys. The inaugural overview of SIM in 2024 approached the topic through the lens of multiple-input multiple-output (MIMO) transceiver design \cite{WC_2024_An_Stacked}, highlighting the potential superiority of SIM over conventional MIMO schemes in terms of signal processing efficiency, and offering numerical results to substantiate the merits of wave-based signal processing in wireless systems. Building on this foundation, Liu \emph{et al.} \cite{arXiv_2024_Hao_Stacked} presented a higher-dimensional treatment of SIM technology, extending its scope to semantic coding while also addressing the channel estimation challenges inherent to SIM-assisted communication systems. A more concise perspective was subsequently offered in \cite{EUCAP_2025_Renzo_State}, which sketched a preliminary overview of prevailing SIM-related applications. More recently, \cite{Electronics_2026_Liu_Stacked} delivered a systematic examination of SIM technology encompassing its multi-layer stacking architectures and wave-domain signal processing principles, and further explored its application prospects across critical 6G use cases, including integrated sensing and communication (ISAC), low Earth orbit (LEO) satellite communications, and unmanned aerial vehicle (UAV) networks. In \cite{arXiv_2026_Sheemar_A}, Sheemar \emph{et al.} offered a comprehensive treatment of SIM from an electromagnetic signal processing standpoint. They critically reviewed existing modeling methodologies grounded in cascaded operator formulations, multi-port impedance representations, and network parameter characterizations, and examined their respective implications for scalable optimization and holistic system design.

In 2026, a special issue on SIM, organized under our editorial leadership in \emph{IEEE Wireless Communications}, culminated in the formal publication of $10$ overview articles, collectively addressing a broad spectrum of SIM-related topics spanning channel modeling, cross-domain integration, distributed deployment, and optimization methodologies, among others. Representative contributions from this special issue are highlighted below. Shi \emph{et al.} \cite{WC_2026_Shi_Stacked} examined the principal signal processing challenges arising from the integration of SIMs into distributed wireless networks, with particular attention to hierarchical framework design, user association strategies, and joint precoding schemes. The accompanying case study further corroborated the substantial performance gains attainable through SIM-enabled network architectures. Sun \emph{et al.} \cite{WC_2026_Sun_Cross} introduced the concept of cross-domain collaborative SIM (C-SIM), a unified framework designed to seamlessly consolidate high-degree-of-freedom (DoF) electromagnetic wave-based digital beamforming, parallel optical semantic computing, and flexible polarization operations, while simultaneously alleviating the hardware overhead and implementation complexity associated with conventional approaches. Illustrative case studies and numerical evaluations were provided to validate the potential of C-SIM in achieving appreciable performance improvements within future 6G systems.
\begin{table*}[!t]
\centering
\renewcommand\arraystretch{1.25}
\caption{A comparison of this survey with existing overview and survey papers on SIM.}
\begin{tabular}{l|c|l}
\hline
 & \begin{tabular}[c]{@{}c@{}}Type\end{tabular} & Primary perspective, modeling framework, hardware coverage, and unique contributions \\ \hline\hline
Shi \emph{et al.} \cite{WC_2026_Shi_Stacked} & Magazine & \begin{tabular}[c]{@{}l@{}}\textbf{\emph{Primary perspective:}} System-level distributed wireless network paradigm via electromagnetic control\\\textbf{\emph{Modeling framework:}} Hierarchical coordination architectures combined with multi-layer wave manipulation\\\textbf{\emph{Hardware coverage:}} Active SIMs, passive transmissive SIMs, and passive transmissive-reflective SIMs\\\textbf{\emph{Unique contribution:}}\\\quad \emph{1.} Explores SIM-aided distributed networks to achieve wide-area coverage and real-time environment shaping\\\quad \emph{2.} Evaluates hierarchical frameworks (centralized, hybrid, and distributed) to manage network signaling overhead\\\quad \emph{3.} Validates bipartite matching for antenna-UE association in joint precoding designs\end{tabular}\\ \hline 
Sun \emph{et al.} \cite{WC_2026_Sun_Cross} & Magazine & \begin{tabular}[c]{@{}l@{}}\textbf{\emph{Primary perspective:}} Cross-domain collaboration (spatial + semantic + polarization)\\\textbf{\emph{Modeling framework:}} Inter-layer near-field model combined with far-field wireless channel for end-to-end C-SIM\\\textbf{\emph{Hardware coverage:}} Passive, active, and hybrid SIM classification with equivalent circuit models\\\textbf{\emph{Unique contribution:}}\\\quad \emph{1.} Proposes the novel C-SIM architecture integrating three domains in one EM layer\\\quad \emph{2.} Introduces dual-polarized SIM for polarization manipulation without RF chains\\\quad \emph{3.} Demonstrates cross-domain AJC-RSMA case study with numerical validation\end{tabular} \\ \hline
Liu \emph{et al.} \cite{Electronics_2026_Liu_Stacked} & Survey & \begin{tabular}[c]{@{}l@{}}\textbf{\emph{Primary perspective:}} Architecture \& 6G scenario adaptation\\\textbf{\emph{Modeling framework:}} Standard cascaded model based on Rayleigh–Sommerfeld equation\\\textbf{\emph{Hardware coverage:}} F4B substrate metalens prototype with alignment tolerance and insertion loss\\\textbf{\emph{Unique contribution:}}\\\quad \emph{1.} Explicit comparison of SIM vs. (holographic) MIMO, active RIS, and phased arrays\\\quad \emph{2.} Discusses hardware prototype challenges, inculding inter-layer alignment, insertion loss, etc.\\\quad \emph{3.} Algorithm comparison table for wave-domain beamforming\end{tabular} \\ \hline
Sheemar \emph{et al.} \cite{arXiv_2026_Sheemar_A} & Survey & \begin{tabular}[c]{@{}l@{}}\textbf{\emph{Primary perspective:}} Electromagnetic \& multiport network theory\\\textbf{\emph{Modeling framework:}} Multiport ECO formulation with mutual coupling, near-field loss, and impedance matching\\\textbf{\emph{Hardware coverage:}} Static, programmable-passive, programmable-active SIMs\\\textbf{\emph{Unique contribution:}}\\\quad \emph{1.} First unified survey grounding SIM in Z-, S-, and T-parameter network theory\\\quad \emph{2.} Rigorous comparative table of all three modeling frameworks with tradeoff analysis\\\quad \emph{3.} Taxonomy of emerging SIM architectures (delay-augmented, dual-polarized, flexible, meta-fiber, nonlinear)\end{tabular} \\ \hline
$\star$ This paper & Survey & \begin{tabular}[c]{@{}l@{}}\textbf{\emph{Primary perspective:}} Wave-domain signal processing \& electromagnetic neural networks\\\textbf{\emph{Modeling framework:}} Simplified diagonal propagation model with focuses on DNN analogy and SIM configuration\\\textbf{\emph{Hardware coverage:}} Includes experimental prototype results for DOA estimation, ISAC, and image classification\\\textbf{\emph{Unique contribution:}}\\\quad \emph{1.} Traces SIM evolution as fusion of neural networks, EM computing, and metasurfaces\\\quad \emph{2.} Introduces antenna selection \& port association as a formal research problem\\\quad \emph{3.} Comprehensive prototype hardware survey on layer count, size, frequency across THz to sub-6 GHz\\\quad \emph{4.} A detailed summary table on optimization variables and solvers\\\quad \emph{5.} Discusses potential propagation coefficient calibration methodology\end{tabular} \\ \hline
\end{tabular}
\label{tab:comparison}\vspace{-0.4cm}
\end{table*}

Notwithstanding the aforementioned contributions, the present survey distinguishes itself from the existing literature across several important dimensions, offering a number of unique perspectives that have not been collectively addressed in prior works. \emph{First}, a more comprehensive account of the developmental history of SIM is presented, tracing its evolution through the lens of technological integration. Such a historical perspective is instrumental in furnishing fresh viewpoints and conceptual insights to guide future research directions. \emph{Second}, a thorough and systematic review of SIM hardware implementations is conducted, spanning the frequency spectrum from sub-6 GHz to terahertz (THz) bands. This review encompasses a detailed characterization of hardware parameters, layer configurations, physical dimensions, functional capabilities, operating frequencies, and reported performance metrics, offering a level of hardware-oriented depth. \emph{Third}, a detailed examination is provided of works in which SIM serves as an optimization parameter within wireless communication system design. This review covers optimization objectives, tunable parameters, representative hardware constraints, and potential mitigation strategies, thereby offering readers both a consolidated reference and a source of inspiration for future system-level investigations. \emph{Fourth}, in-depth treatments of semantic encoding, parameter estimation, imaging, and logical operations are presented, and these functionalities are systematically incorporated into a unified framework encompassing communication, sensing, and computing. This integrative perspective is intended to foster a more holistic understanding of the broader potential of SIM and electromagnetic-domain signal processing, and to stimulate further interdisciplinary research. \emph{Fifth}, several critical challenges pertaining to the practical deployment of SIM are reviewed, including channel estimation and propagation modeling. Notably, for the first time in the literature, the selection and association of ports at both ends of the SIM are formally introduced, and a principled solution to this problem is delineated. In addition, the estimation of propagation parameters under conditions of SIM hardware mismatch is rigorously examined. \emph{Sixth}, a comprehensive review of strategies for enhancing SIM capabilities is conducted. This includes approaches to augmenting the number of available channels from the perspectives of spatial planning, frequency diversity, and orbital angular momentum (OAM), as well as the introduction of nonlinearity at the device level to enhance the network's inferential capacity. For clarity and ease of reference, a structured comparison between the present survey and the related literature is summarized in Table~\ref{tab:comparison}.

\section{Electromagnetic Foundation of SIM}
From a physical perspective, a SIM consists of multiple layers of reconfigurable transmissive metasurfaces arranged in a stacked configuration. Grounded in the Huygens–Fresnel principle~\cite{NE_2022_Liu_A}, each meta-atom within a given metasurface layer acts as a secondary source — analogous to a point emitter — radiating a spherical wave that subsequently illuminates all meta-atoms on the immediately succeeding layer. The phase and amplitude of the resulting diffracted wave arriving at the next layer are collectively governed by three factors: the electromagnetic waves incident upon the meta-atoms of the current layer, the complex-valued transmission coefficients characterizing that layer, and the complex-valued free-space propagation coefficients between the two layers under consideration~\cite{WC_2024_An_Stacked, arXiv_2024_Hao_Stacked}. Based on the Rayleigh–Sommerfeld diffraction integral~\cite{Sci_2018_Lin_All, BOOK_2013_Born_Principles}, the propagation coefficient between the $\tilde{n}$-th meta-atom on the $\left ( l-1 \right )$-st metasurface layer and the $n$-th meta-atom on the $l$-th metasurface layer is given by\footnote{The propagation model in \eqref{eq_1} allows us to account for near-field propagation between adjacent metasurface layers, but it can be applied provided that some physical conditions are fulfilled. Notably, the (sub-wavelength) inter-layer distance cannot be too small, so as to ensure that \eqref{eq_1} is physically consistent. As a rule of thumb, the inter-layer distance, given the size of meta-atoms, should be sufficiently large to ensure that the amplitude of \eqref{eq_1} is less than one \cite{OE_2017_Mehrabkhani_Is}.}
\begin{align}\label{eq_1}
 w_{n,\tilde{n}}^{l}=\frac{\mathcal{A} \cos \zeta _{n,\tilde{n}}^{l}}{d_{n,\tilde{n}}^{l}}\left ( \frac{1}{2\pi d_{n,\tilde{n}}^{l}} -j\frac{1}{\lambda }\right )e^{j2\pi d_{n,\tilde{n}}^{l}/\lambda },
\end{align}
where $d_{n,\tilde{n}}^{l}$ denotes the corresponding transmission distance. Furthermore, $\mathcal{A}$ is the area of each meta-atom, $\zeta _{n,\tilde{n}}^{l}$ represents the angle between the propagation direction and the normal direction of the $\left ( l-1 \right )$-st metasurface layer, and $\lambda$ represents the radio wavelength. The total electromagnetic field that impinges upon each meta-atom on a metasurface layer is the sum of the fields refracted by all the meta-atoms on the previous layer~\cite{JSAC_2022_An_Stacked}.

Architecturally, a SIM mirrors a conventional ANN, comprising an input layer, one or more hidden layers, and an output layer \cite{NE_2022_Liu_A, EUCAP_2025_Renzo_State}. Within this framework, each hidden layer is realized by a transmissive metasurface—a dense array of electronically programmable meta-atoms that function analogously to artificial neurons. Inter-layer connectivity is established via the spatial diffraction of electromagnetic waves, meaning that, in principle, any given meta-atom can couple with any other across the structure. To ensure a predictable forward propagation, however, practical SIM implementations utilize anti-reflective coatings to suppress higher-order internal reflections and mitigate coupling between non-adjacent layers \cite{LSA_2022_Luo_Metasurface, Access_2026_Arya_A}. Consequently, the SIM operates equivalently to a feed-forward diffractive network where waves propagate monotonically from one layer to the next. Crucially, whereas software-defined ANNs optimize abstract weights mapping successive layers alongside layer-specific biases, the complex-valued propagation coefficients between metasurface layers in a SIM are strictly dictated by the physical wave propagation direction and axial inter-layer spacing, rendering them unalterable \cite{Sci_2018_Lin_All}. Instead, the trainable DoFs in a SIM-based diffractive network are confined strictly to the transmission coefficients of the individual meta-atoms at their specific lattice coordinates \cite{APR_2024_Chen_Diffractive}.

\begin{figure}[!t]
\centering
\includegraphics[width=8.5cm]{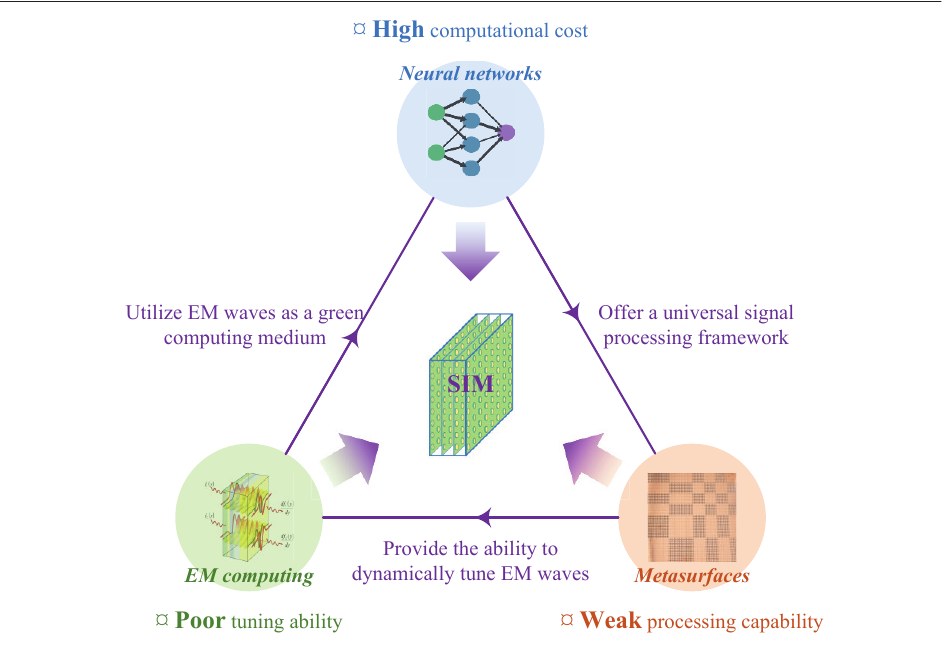}
\caption{SIM is the amalgamation of neural networks, electromagnetic computing, and metasurfaces. The advantages of these three technologies complement each other.}
\label{fig_3}\vspace{-0.4cm}
\end{figure}

\begin{table*}[!t]
\centering
\scriptsize
\renewcommand\arraystretch{1.25}
\caption{A survey of existing SIM prototypes.}
\label{tab1}
\begin{tabular}{l|l|l|l|l|l|l}
\hline
\begin{tabular}[c]{@{}l@{}}$\sharp$ Reference\\$\flat$ Author\\$\natural$ Year\end{tabular} & Prototype & \begin{tabular}[c]{@{}l@{}}$\diamondsuit$ SIM architecture\\$\clubsuit$ Per-layer size\\$\heartsuit$ Inter-layer spacing\\$\spadesuit$ Phase modulation\end{tabular} & \begin{tabular}[c]{@{}l@{}}$\circledast$ Task\\$\circleddash$ Loss function\\$\circledcirc$ Dataset\end{tabular} & \begin{tabular}[c]{@{}l@{}}$f$: Frequency\\$\lambda$: Wavelength\end{tabular} & Output & Performance \\ \hline\hline
\multirow{3}{*}{\begin{tabular}[c]{@{}l@{}}\\ \\ \\$\sharp$~\cite{Sci_2018_Lin_All}\\$\flat$ Lin \emph{et al.}\\$\natural$ 2018\end{tabular}} & \makecell[c]{\multirow{2}{*}{\raisebox{-0\height}{\includegraphics[width=1.6cm]{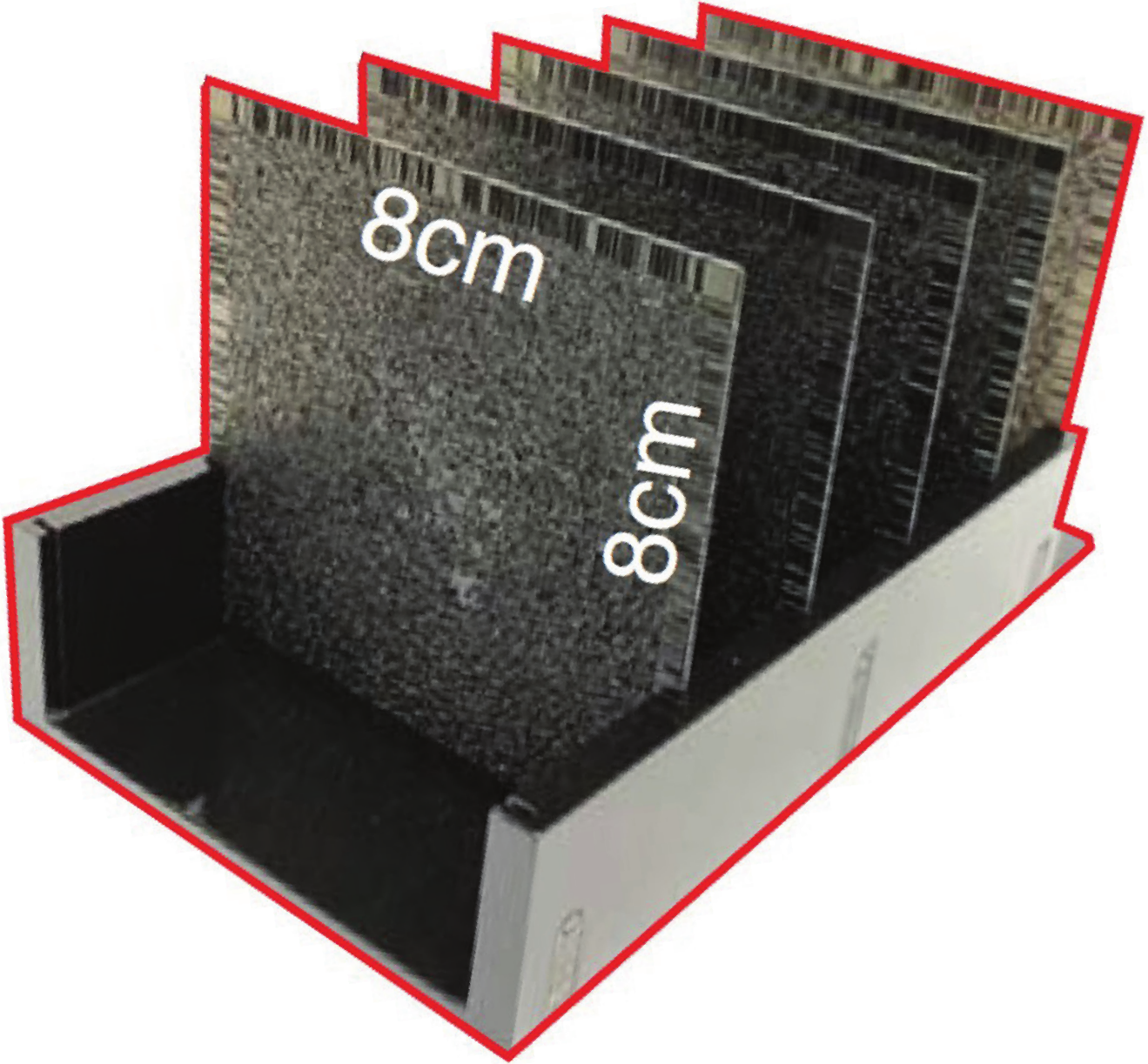}}}} & \multirow{2}{*}{\begin{tabular}[c]{@{}l@{}}$\diamondsuit$ $200 \times 200 \times 5$\\$\clubsuit$ $8$ cm $\times$ $8$ cm\\$\heartsuit$ $3$ cm\\$\spadesuit$ $\left [ 0,\pi \right )$\end{tabular}} & \begin{tabular}[c]{@{}l@{}}$\circledast$ Image classification\\$\circleddash$ Energy distribution\\$\circledcirc$ MNIST\end{tabular} & \multirow{3}{*}{\begin{tabular}[c]{@{}l@{}}\\ \\ \\ \\$f$: $0.40$ THz\\$\lambda$: $0.75$ mm\end{tabular}} & \multirow{2}{*}{\begin{tabular}[c]{@{}l@{}}\\$10$ detection regions\end{tabular}} & \begin{tabular}[c]{@{}l@{}}Classification accuracy\\$91.75\%$ (Simulation)\\$80.74\%$ (Experiment)\end{tabular} \\ \cline{4-4} \cline{7-7} 
 & & & \begin{tabular}[c]{@{}l@{}}$\circledast$ Image classification\\$\circleddash$ Energy distribution\\$\circledcirc$ Fashion-MNIST\end{tabular} & & & \begin{tabular}[c]{@{}l@{}}Classification accuracy\\$81.13\%$ (Simulation)\\$73.02\%$ (Experiment)\end{tabular} \\ \cline{2-4} \cline{6-7} 
 & \makecell[c]{\raisebox{-0.9\height}{\includegraphics[height=1.3cm]{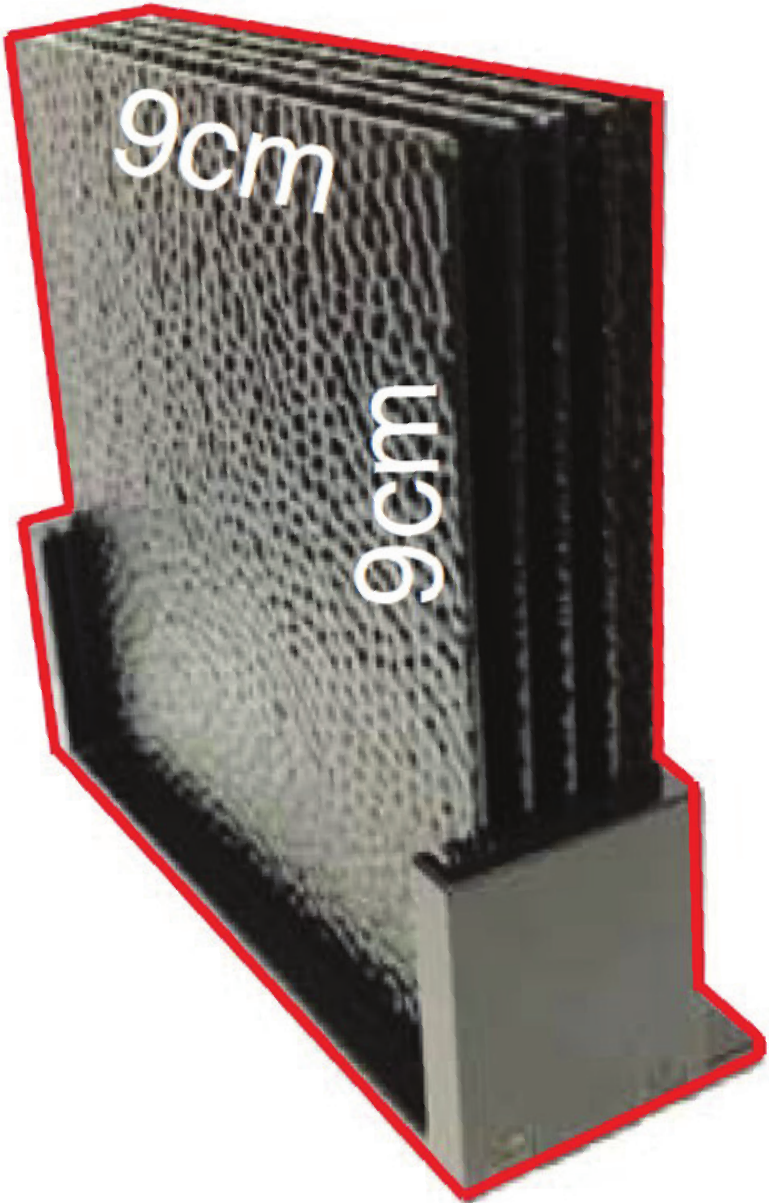}}} & \begin{tabular}[c]{@{}l@{}}$\diamondsuit$ $300 \times 300 \times 5$\\$\clubsuit$ $9$ cm $\times$ $9$ cm\\$\heartsuit$ $4$ cm\\$\spadesuit$ $\left [ 0,2\pi \right )$\end{tabular} & \begin{tabular}[c]{@{}l@{}}$\circledast$ Amplitude imaging\\$\circleddash$ MSE\\$\circledcirc$ ImageNet\end{tabular} & &Plane: $9$ cm $\times$ $9$ cm& \begin{tabular}[c]{@{}l@{}}Resolution\\$1.2$ mm (Simulation)\\$1.8$ mm (Experiment)\end{tabular} \\ \hline
\begin{tabular}[c]{@{}l@{}}$\sharp$~\cite{LSA_2020_Qian_Performing}\\$\flat$ Qian \emph{et al.}\\$\natural$ 2020\end{tabular} & \makecell[c]{\raisebox{-0.8\height}{\includegraphics[height=1.3cm]{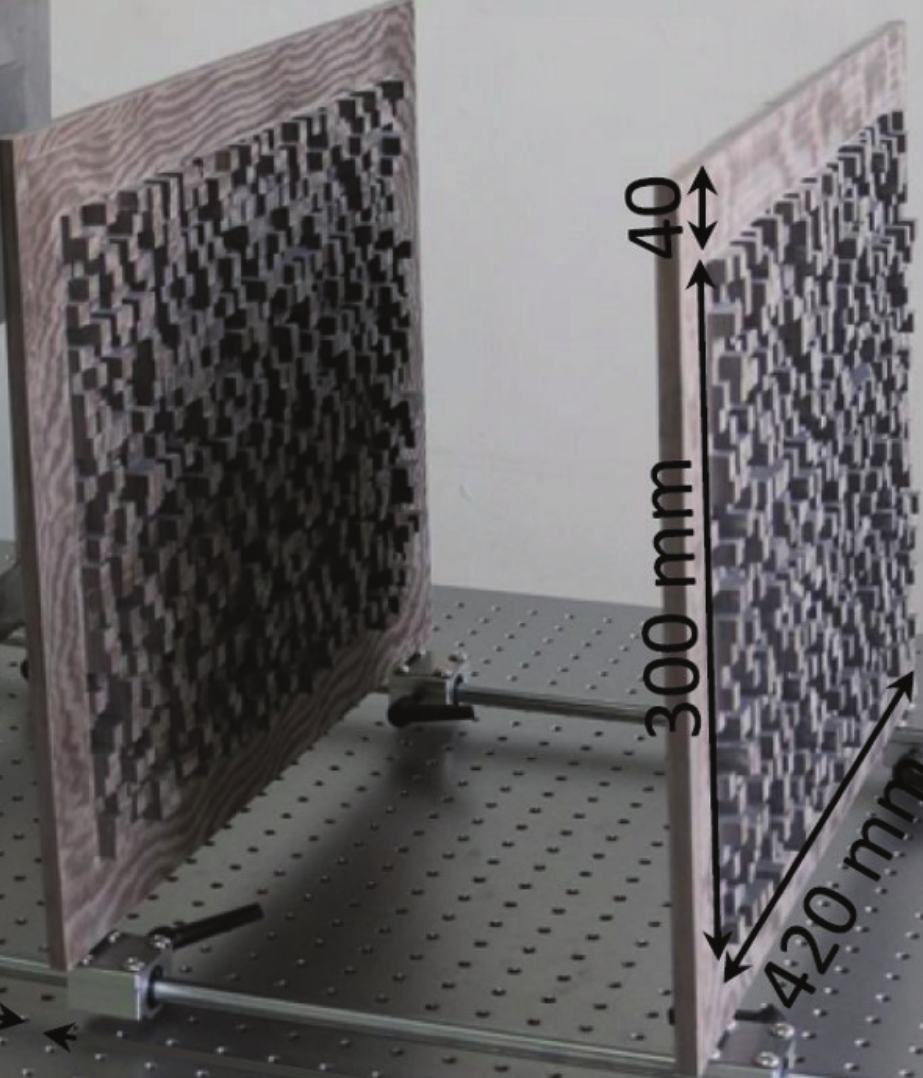}}} & \begin{tabular}[c]{@{}l@{}}$\diamondsuit$ $30 \times 42 \times 2$\\$\clubsuit$ $30$ cm $\times$ $42$ cm\\$\heartsuit$ $30$ cm\\$\spadesuit$ $\left [ 0,2\pi \right )$\end{tabular} & \begin{tabular}[c]{@{}l@{}}$\circledast$ Logic operation\\$\circleddash$ MSE\\$\circledcirc$ -- -- --\end{tabular} & \begin{tabular}[c]{@{}l@{}}$f$: $17$ GHz\\$\lambda$: $1.76$ mm\end{tabular} & $2$ detection regions &\begin{tabular}[c]{@{}l@{}}Contrast ratio\\$> 9.6$ dB\end{tabular} \\ \hline
\multirow{4}{*}{\begin{tabular}[c]{@{}l@{}}\\ \\$\sharp$~\cite{SA_2021_Li_Spectrally}\\$\flat$ Li \emph{et al.}\\$\natural$ 2021\end{tabular}} & \makecell[c]{\multirow{4}{*}{\raisebox{-1.45\height}{\includegraphics[width=1.6cm]{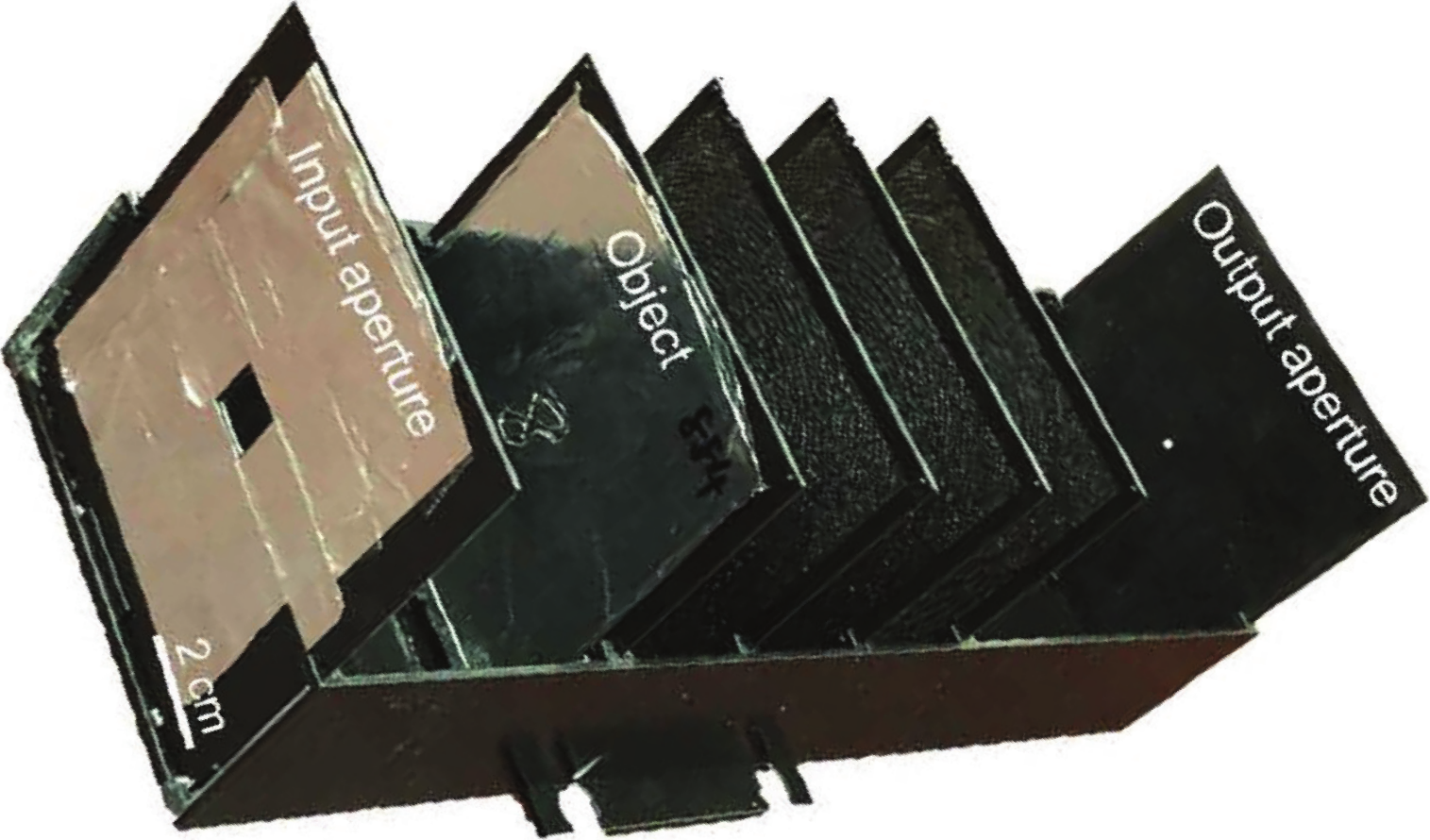}}}} & \multirow{4}{*}{\begin{tabular}[c]{@{}l@{}}\\$\diamondsuit$ $160 \times 160 \times 3$\\$\clubsuit$ $8$ cm $\times$ $8$ cm\\$\heartsuit$ $3$ cm\\$\spadesuit$ $12$ bits\end{tabular}} & \multirow{2}{*}{\begin{tabular}[c]{@{}l@{}}$\circledast$ Image classification\\$\circleddash$ Softmax CE error\\$\circledcirc$ MNIST\end{tabular}} & \begin{tabular}[c]{@{}l@{}}$f$: $0.21$ THz $\sim$ $0.30$ THz\\$\lambda$: $1.00$ mm $\sim$ $1.45$ mm\end{tabular} & \begin{tabular}[c]{@{}l@{}}$1$ broadband detector\\with $10$ wavelengths\end{tabular} &\begin{tabular}[c]{@{}l@{}}Classification accuracy\\$96.01\%$\end{tabular}\\ \cline{5-7} 
 & & & & \begin{tabular}[c]{@{}l@{}}$f$: $0.19$ THz $\sim$ $0.46$ THz\\$\lambda$: $0.65$ mm $\sim$ $1.60$ mm\end{tabular} & \begin{tabular}[c]{@{}l@{}}$1$ broadband detector\\with $20$ wavelengths\end{tabular} &\begin{tabular}[c]{@{}l@{}}Classification accuracy\\$96.04\%$\end{tabular}\\ \cline{4-7} 
 & & & \multirow{2}{*}{\begin{tabular}[c]{@{}l@{}}$\circledast$ Image classification\\$\circleddash$ Softmax CE error\\$\circledcirc$ EMNIST\end{tabular}} & \begin{tabular}[c]{@{}l@{}}$f$: $0.21$ THz $\sim$ $0.36$ THz\\$\lambda$: $0.83$ mm $\sim$ $1.45$ mm\end{tabular} & \begin{tabular}[c]{@{}l@{}}$1$ broadband detector\\with $26$ wavelengths\end{tabular} &\begin{tabular}[c]{@{}l@{}}Classification accuracy\\$84.05\%$\end{tabular}\\ \cline{5-7} 
 & & & & \begin{tabular}[c]{@{}l@{}}$f$: $0.20$ THz $\sim$ $0.40$ THz\\$\lambda$: $0.76$ mm $\sim$ $1.52$ mm\end{tabular} & \begin{tabular}[c]{@{}l@{}}$1$ broadband detector\\with $52$ wavelengths\end{tabular} &\begin{tabular}[c]{@{}l@{}}Classification accuracy\\$85.60\%$\end{tabular}\\ \hline
\multirow{4}{*}{\begin{tabular}[c]{@{}l@{}}\\ \\ \\$\sharp$~\cite{NE_2022_Liu_A}\\$\flat$ Liu \emph{et al.}\\$\natural$ 2022\end{tabular}} & \makecell[c]{\multirow{4}{*}{\raisebox{-1.45\height}{\includegraphics[width=1.6cm]{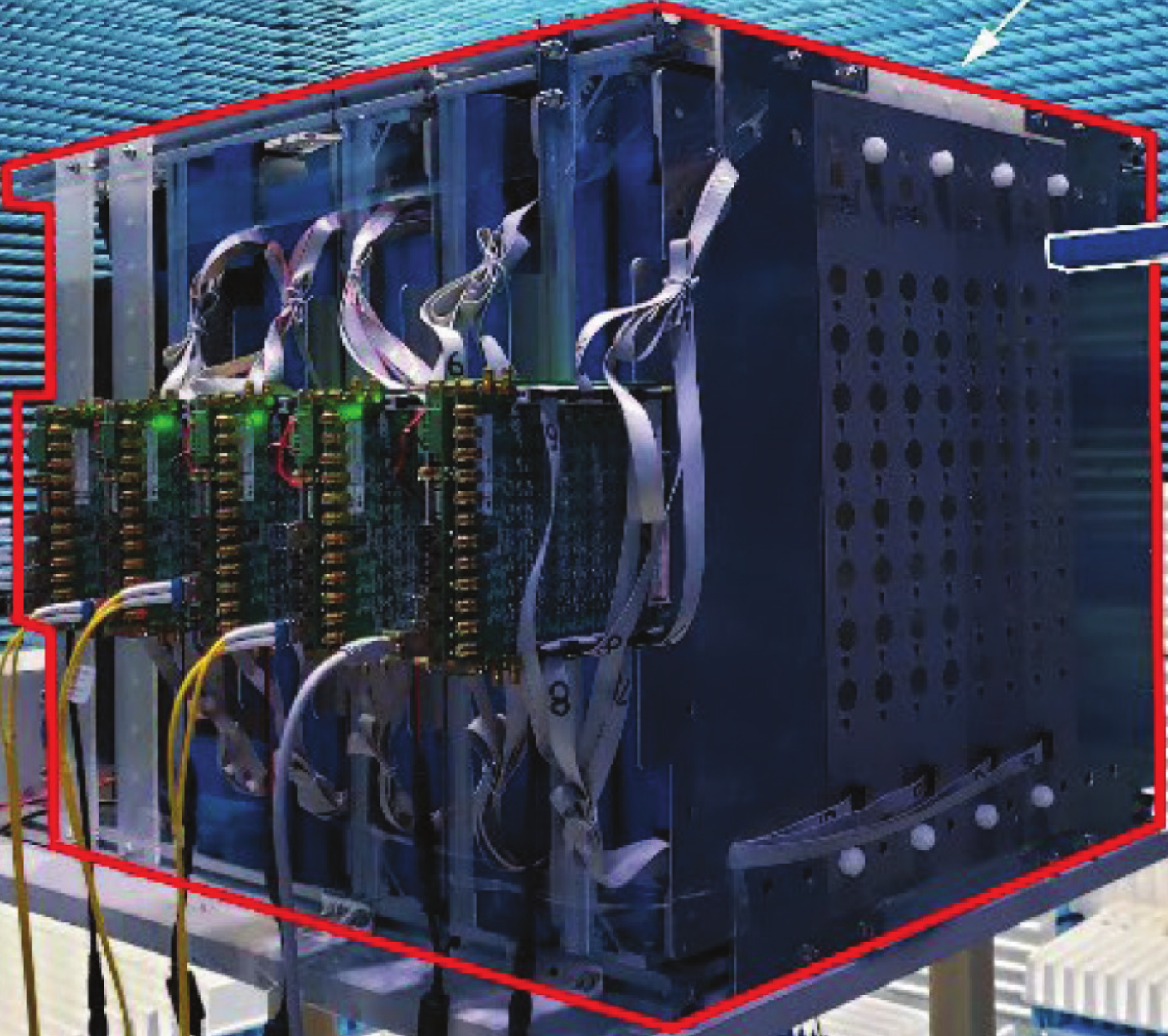}}}} & \multirow{4}{*}{\begin{tabular}[c]{@{}l@{}}\\ \\ \\$\diamondsuit$ $8 \times 8 \times 5$\\$\clubsuit$ $56$ cm $\times$ $56$ cm\\$\heartsuit$ $10$ cm\\$\spadesuit$ $4$ bits\end{tabular}} & \begin{tabular}[c]{@{}l@{}}$\circledast$ Image classification\\$\circleddash$ CE error\\$\circledcirc$ `I' and `{[}{]}'\end{tabular} & \multirow{2}{*}{\begin{tabular}[c]{@{}l@{}}\\$f$: $5.40$ GHz\\$\lambda$: $5.56$ cm\end{tabular}} & $2$ detection regions & \begin{tabular}[c]{@{}l@{}}Classification accuracy\\$100\%$\end{tabular}\\ \cline{4-4} \cline{6-7} 
 & & & \begin{tabular}[c]{@{}l@{}}$\circledast$ Image classification\\$\circleddash$ CE error\\$\circledcirc$ MNIST\end{tabular} & & $10$ detection regions &\begin{tabular}[c]{@{}l@{}}Classification accuracy\\$90.76\%$\end{tabular}\\ \cline{4-7} 
 & & & \begin{tabular}[c]{@{}l@{}}$\circledast$ CDMA\\$\circleddash$ MSE\\$\circledcirc$ -- -- --\end{tabular} & \multirow{2}{*}{\begin{tabular}[c]{@{}l@{}}\\$f$: $5.50$ GHz\\$\lambda$: $5.45$ cm\end{tabular}} & $4$ detection regions &\begin{tabular}[c]{@{}l@{}}Bit error rate\\$5.2\times10^{-3}$\end{tabular}\\ \cline{4-4} \cline{6-7} 
 & & & \begin{tabular}[c]{@{}l@{}}$\circledast$ Multi-beam focusing\\$\circleddash$ MSE\\$\circledcirc$ -- -- --\end{tabular} & & $2$ target points & \begin{tabular}[c]{@{}l@{}}Energy leakage\\$<10\%$\end{tabular}\\ \hline
 \begin{tabular}[c]{@{}l@{}}$\sharp$~\cite{AOM_2024_Gu_Classification}\\$\flat$ Gu \emph{et al.}\\$\natural$ 2024\end{tabular} & \makecell[c]{\raisebox{-0.8\height}{\includegraphics[width=1.6cm]{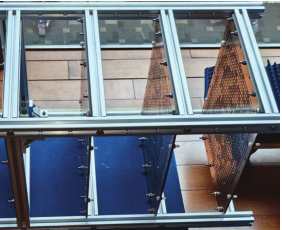}}}& \begin{tabular}[c]{@{}l@{}}$\diamondsuit$ $32 \times 32 \times 3$\\$\clubsuit$ $34$ cm $\times$ $34$ cm\\$\heartsuit$ $18$ cm\\$\spadesuit$ $\left [ 0,2\pi \right )$\end{tabular} & \begin{tabular}[c]{@{}l@{}}$\circledast$ Image classification\\$\circleddash$ CE error\\$\circledcirc$ MNIST\end{tabular} & \begin{tabular}[c]{@{}l@{}}$f$: $11$ GHz\\$\lambda$: $2.73$ cm\end{tabular} & $10$ detection regions &\begin{tabular}[c]{@{}l@{}}Classification accuracy\\$90\%$\end{tabular} \\ \hline
\begin{tabular}[c]{@{}l@{}}$\sharp$~\cite{Arxiv_2024_Wang_Multi}\\$\flat$ Wang \emph{et al.}\\$\natural$ 2024\end{tabular} & \makecell[c]{\raisebox{-0.8\height}{\includegraphics[height=1.3cm]{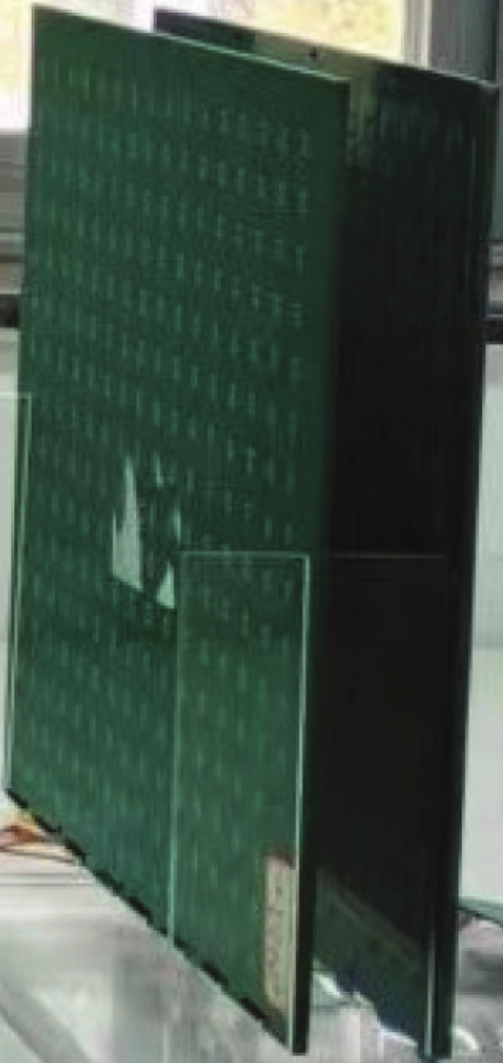}}}& \begin{tabular}[c]{@{}l@{}}$\diamondsuit$ $16 \times 16 \times \left ( 1\sim 3 \right )$\\$\clubsuit$ $31$ cm $\times$ $31$ cm\\$\heartsuit$ $1.29$ cm $\sim$ $10.34$ cm\\$\spadesuit$ $1$ bit\end{tabular} & \begin{tabular}[c]{@{}l@{}}$\circledast$ Signal enhancement\\$\circleddash$ Received power\\$\circledcirc$ -- -- --\end{tabular} & \begin{tabular}[c]{@{}l@{}}$f$: $5.80$ GHz\\$\lambda$: $5.17$ cm\end{tabular} & $1$ receiving antenna &\begin{tabular}[c]{@{}l@{}}Power gain\\$7$ dB\end{tabular} \\ \hline
\multirow{2}{*}{\begin{tabular}[c]{@{}l@{}}\\$\sharp$~\cite{LSA_2024_Gao_Super}\\$\flat$ Gao \emph{et al.}\\$\natural$ 2024\end{tabular}} & \makecell[c]{\raisebox{-0.8\height}{\includegraphics[height=1.3cm]{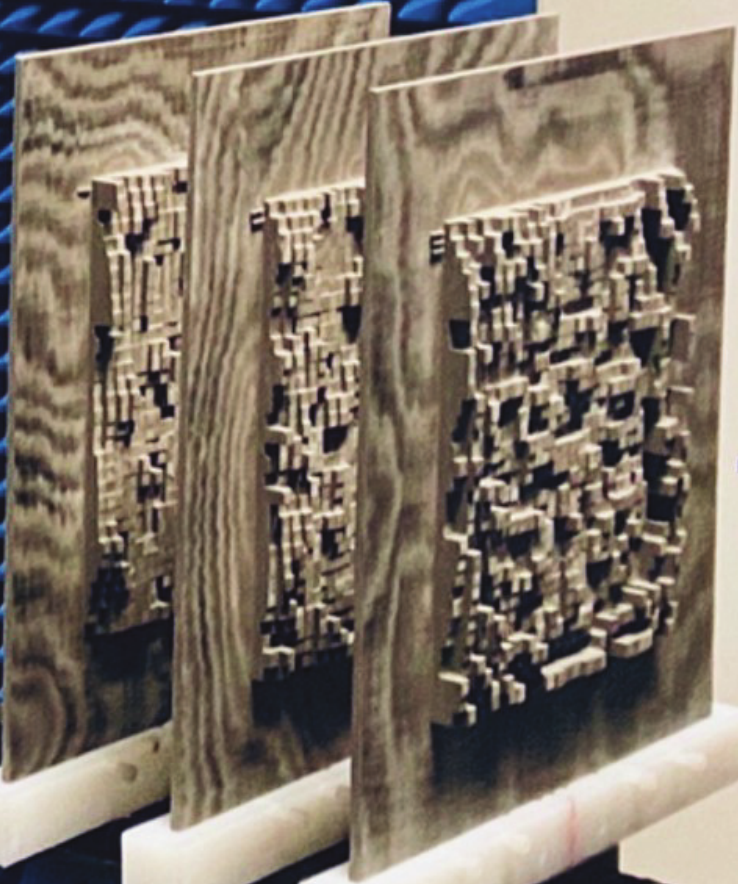}}}& \begin{tabular}[c]{@{}l@{}}$\diamondsuit$ $32 \times 32 \times 3$\\$\clubsuit$ $27.45$ cm $\times$ $27.45$ cm\\$\heartsuit$ $5.45$ cm\\$\spadesuit$ $7$ bits\end{tabular} & \multirow{2}{*}{\begin{tabular}[c]{@{}l@{}}\\$\circledast$ DOA estimation\\$\circleddash$ MSE \& CE error\\$\circledcirc$ -- -- --\end{tabular}} & \multirow{2}{*}{\begin{tabular}[c]{@{}l@{}}\\ \\$f$: $25$ GHz $\sim$ $30$ GHz\\$\lambda$: $1.0$ cm $\sim$ $1.2$ cm\end{tabular}} & \multirow{2}{*}{\begin{tabular}[c]{@{}l@{}}\\ \\$10$ detection regions\end{tabular}} &\begin{tabular}[c]{@{}l@{}}Angular resolution\\$3^{\circ}$ in $\left [ -15^{\circ},15^{\circ} \right ]$\end{tabular}\\ \cline{2-3} \cline{7-7} 
 & \makecell[c]{\raisebox{-0.9\height}{\includegraphics[height=1.3cm]{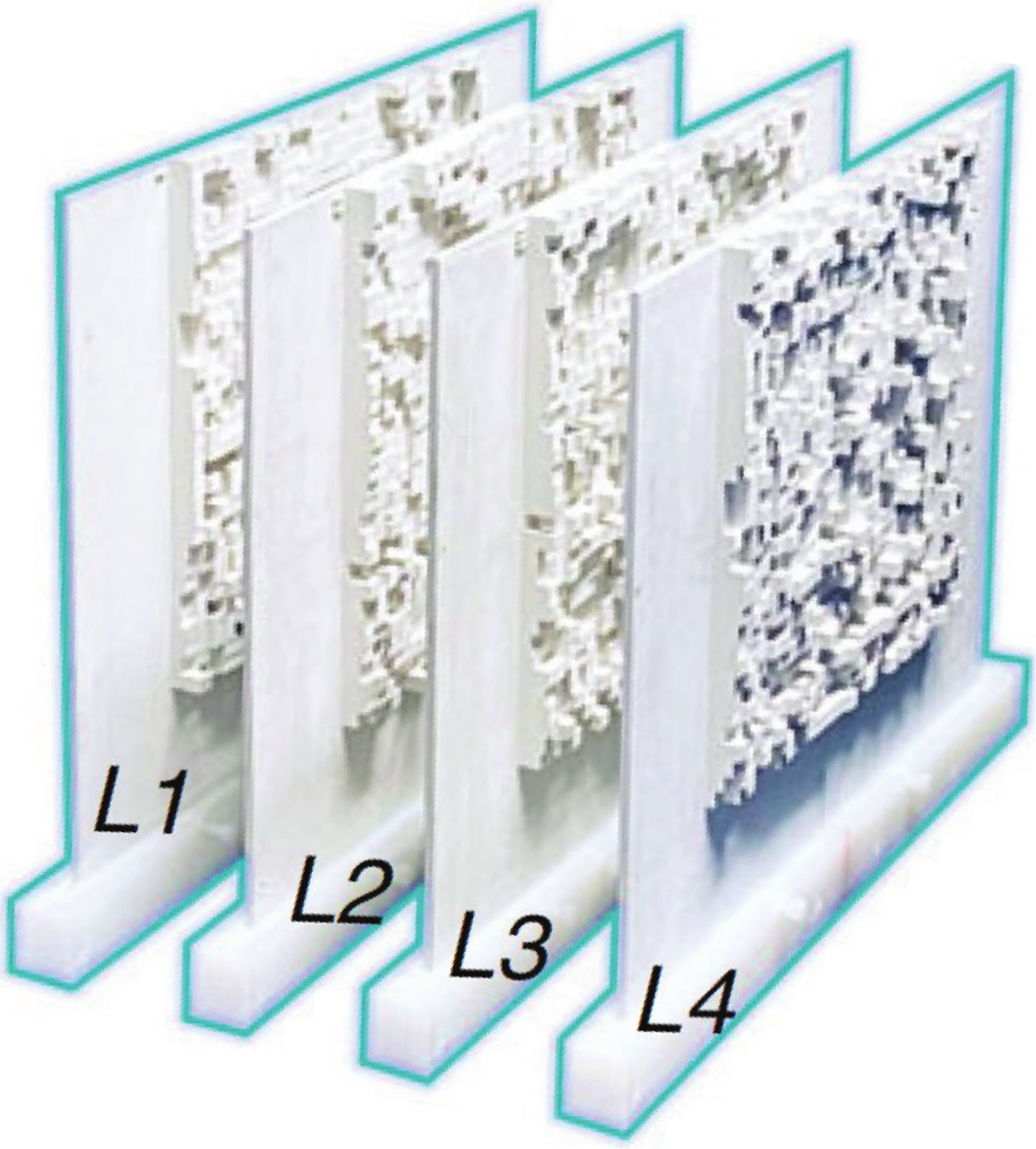}}} & \begin{tabular}[c]{@{}l@{}}$\diamondsuit$ $32 \times 32 \times 4$\\$\clubsuit$ $27.45$ cm $\times$ $27.45$ cm\\$\heartsuit$ $5.45$ cm\\$\spadesuit$ $7$ bits\end{tabular} & & & &\begin{tabular}[c]{@{}l@{}}Angular resolution\\$1^{\circ}$ in $\left [ -5^{\circ},5^{\circ} \right ]$\end{tabular}\\ \hline
\begin{tabular}[c]{@{}l@{}}$\sharp$~\cite{NC_2025_Guo_Polarization}\\$\flat$ Guo \emph{et al.}\\$\natural$ 2025\end{tabular} & \makecell[c]{\raisebox{-0.8\height}{\includegraphics[height=1.3cm]{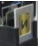}}}& \begin{tabular}[c]{@{}l@{}}$\diamondsuit$ $100 \times 100 \times 2$\\$\clubsuit$ $2.5$ cm $\times$ $2.5$ cm\\$\heartsuit$ $1$ cm\\$\spadesuit$ $\left [ 0,2\pi \right )$\end{tabular} & \begin{tabular}[c]{@{}l@{}}$\circledast$ Image classification\\$\circleddash$ MSE\\$\circledcirc$ EMNIST\end{tabular} & \begin{tabular}[c]{@{}l@{}}$f$: $0.60$ THz\\$\lambda$: $0.50$ mm\end{tabular} & $4$ detection regions &\begin{tabular}[c]{@{}l@{}}Classification accuracy\\$92.50\%$\end{tabular} \\ \hline
\end{tabular}\vspace{-0.4cm}
\end{table*}

A SIM fundamentally represents a synergistic integration of three interdisciplinary technologies: neural networks, electromagnetic computing, and information metasurfaces~\cite{arXiv_2024_Jiancheng_Emerging}. As illustrated in Fig.~\ref{fig_3}, these three cutting-edge technologies exhibit complementary strengths that address each other's inherent limitations. Conventional neural networks, while powerful, depend on digital chips that incur substantial computational overhead and energy consumption~\cite{Proc_2017_Sze_Efficient}. Metasurfaces, despite their sophisticated wavefront manipulation capabilities, possess limited signal processing functionality~\cite{LSA_2014_Cui_Coding}. Similarly, existing electromagnetic computing paradigms suffer from constrained tunability and poor scalability~\cite{Sci_2014_Silva_Performing}. A SIM circumvents these limitations by synthesizing the distinct advantages of all three constituent technologies. Specifically, electromagnetic waves serve as a `green' - namely power-efficient - computational medium for executing forward-propagation inference in neural networks, while neural network architectures provide a versatile framework for signal processing. Concurrently, information metasurfaces enable dynamic modulation of electromagnetic waves, completing this potent tripartite amalgam.

Once the programmable metasurfaces are properly configured, the SIM becomes capable of processing spatial electromagnetic waves as they propagate through the layered structure~\cite{JSAC_2024_An_Two}. Since a SIM directly processes information-carrying electromagnetic waves in free space, it can execute desired computation tasks without requiring any digital storage, transmission, pre-processing of information, as well as external computing power~\cite{arXiv_2024_Jiancheng_Emerging}. Most remarkably, information processing within a SIM takes place at the speed of light~\cite{JSAC_2022_An_Stacked}. Additionally, a SIM can be readily scaled up to accommodate extremely large-scale inputs and a plethora of connections in a cost-effective and power-efficient manner~\cite{NE_2022_Liu_A}. Thanks to these distinctive advantages, several advanced SIMs have been designed to carry out various tasks in the wave domain~\cite{Sci_2018_Lin_All, LSA_2020_Qian_Performing, SA_2021_Li_Spectrally, NE_2022_Liu_A, AOM_2024_Gu_Classification, Arxiv_2024_Wang_Multi, LSA_2024_Gao_Super}. To provide a clearer picture of the hardware landscape, Table \ref{tab1} summarizes state-of-the-art SIM prototypes, meticulously mapping their operational tasks, structural topologies, core hardware parameters, operating frequencies, and empirical performance metrics to offer practical design guidelines for future deployments.

\section{Configuration of SIMs}
As previously underscored, meticulously configuring a SIM is paramount to successfully executing its intended wave-domain functionalities and fulfilling specific inference tasks \cite{arXiv_2024_Jiancheng_Emerging}. To this end, this section elaborates on the optimization and training methodologies conceived for configuring SIM transmission coefficients, analyzing them through two distinct lenses: the functional role assigned to the SIM and the choice between online versus offline training paradigms. Collectively, these paradigms provide a foundational framework for understanding and evaluating the operational behavior of SIMs in practical, real-world deployments.

\begin{figure}[!t]
\centering
\includegraphics[width=8.5cm]{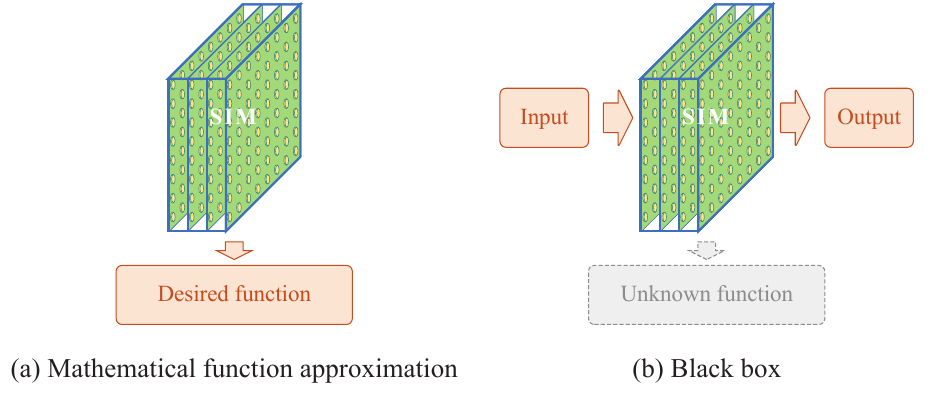}
\caption{Two SIM configuration methods depending on whether the desired transfer function is known or not.}
\label{fig_4}
\end{figure}
\begin{figure}[!t]
\centering
\includegraphics[width=8.5cm]{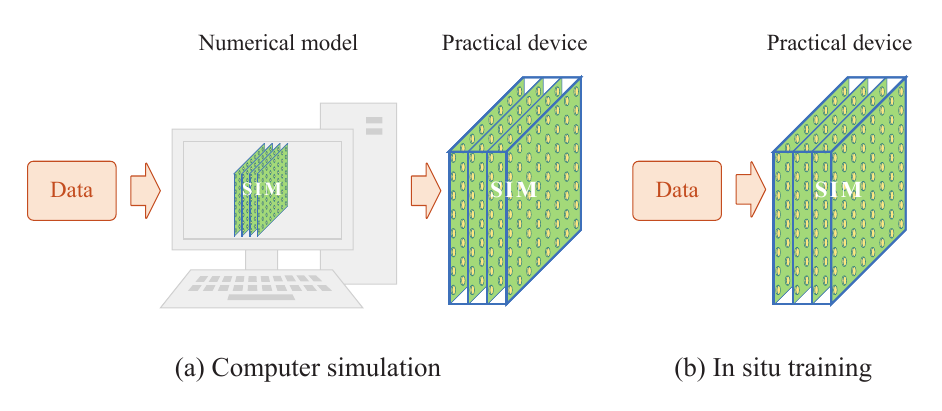}
\caption{Two SIM configuration methods depending on whether there is an accurate numerical model or not.}
\label{fig_5}\vspace{-0.4cm}
\end{figure}
\subsection{Roles of a SIM}
As shown in Fig.~\ref{fig_4}, one can distinguish two primary roles for a SIM according to whether the transformation desired is known or unknown.
\begin{itemize}
 \item \textbf{Mathematical Function Approximation:} In certain application scenarios, the desired transformation matrix -- such as the discrete Fourier transform (DFT) used in spectral analysis -- is known in advance~\cite{JSAC_2024_An_Two}. In these cases, a SIM can be employed to approximate the desired function by tuning the transmission coefficients of multiple metasurface layers. While a digital signal processor can achieve the same function, the utilization of a SIM allows one to achieve near-instantaneous computing, thanks to its electromagnetic-domain signal processing~\cite{JSAC_2022_An_Stacked}. In general, the approximation of desired functions imposes lower requirements on reconfigurability. Therefore, the SIM may only have to be reconfigured on a relatively long timescale~\cite{arXiv_2024_Jiancheng_Emerging,JSAC_2024_An_Two}.
 \item \textbf{Black Box:} In most practical applications, however, such as image classification, the target transformation is often unknown~\cite{LSA_2022_Luo_Metasurface}. In these scenarios, a SIM can be optimized as a ``black box.'' By feeding it with a large number of training samples, the complex-valued transmission coefficients within the SIM can be iteratively adjusted based on their gradients with respect to a custom-designed loss function tailored to the specific wave-domain signal processing task, such as imaging or matrix operations~\cite{SA_2021_Li_Spectrally}. Since SIMs are physically assembled subject to diverse fabrication constraints imposed on phase and amplitude modulation, the transmission coefficients have to be quantized to feasible values that can be practically implemented by individual meta-atoms~\cite{NE_2022_Liu_A}.
\end{itemize}

\begin{figure*}[!t]
\centering
\includegraphics[width=18cm]{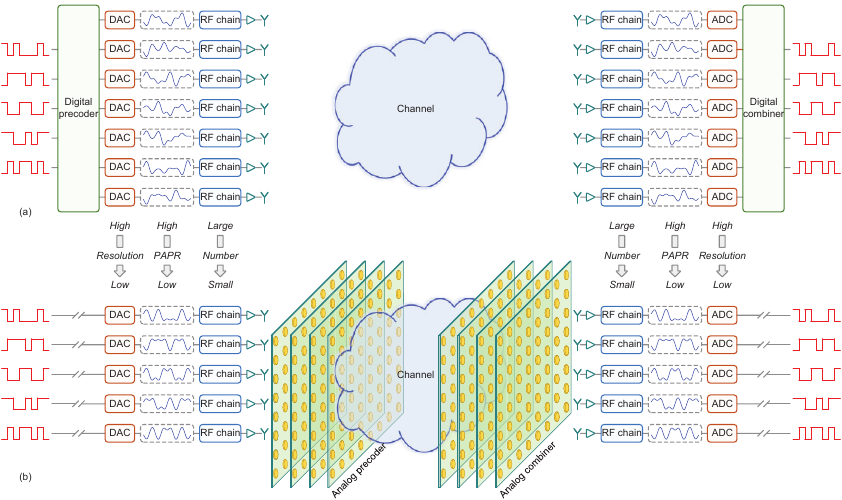}
\caption{A comparison between (a) conventional MIMO architecture and (b) SIM-based MIMO architecture.}
\label{fig_6}\vspace{-0.4cm}
\end{figure*}

\subsection{Training Methods} 
From the perspective of the training process, the transmission coefficients can be trained using either computer simulations or directly on a physical SIM device, as indicated in Fig.~\ref{fig_5}.
\begin{itemize}
\item \textbf{Computer Simulation:} Similar to an ANN, the training of a SIM can be executed on a computer. Once the transmission coefficients are trained to achieve a satisfactory level of approximation accuracy or inference performance, they are deployed on the SIM~\cite{Sci_2018_Lin_All}. Although the training time is constrained by the processing power of the digital computer, the forward propagation during the inference stage occurs at the speed of light. However, computer simulations also pose significant challenges, as they require a reliable mathematical model or simulation tool that can instantly and accurately quantify the complex interactions of wave propagation within the SIM. While established numerical methods, such as the finite-difference time-domain (FDTD) method and finite-elements method (FEM), offer high accuracy, there is still potential for improving their simulation speed~\cite{LSA_2024_Gao_Super}.
\item \textbf{In Situ Training:} In practice, hardware implementations of metasurfaces may diverge from their numerical models due to deformations that occur during the manufacturing process and undesired misalignments between layers. This may lead to inaccurate inference outcomes. To address this issue, an alternative approach known as in situ training has shown considerable potential~\cite{PR_2020_Zhou_In, NMI_2023_Zheng_Dual}. Since the training and inference processes share the same electromagnetic responses, any fabrication imperfections and mechanical misalignments are implicitly accounted for during the training phase, resulting in minimal impact on the inference results. Additionally, the training time is substantially reduced, because the forward propagation of signals occurs at the speed of light~\cite{NE_2022_Liu_A}. Moreover, in situ training enables the application of deep reinforcement learning (DRL) strategies \cite{Nature_2025_Momeni_Training}, which can train the transmission coefficients based on practical observations in a dynamic environment. However, in situ training needs real-time feedback from the output, which introduces extra signaling overhead when the SIM is deployed far from the inference receiver~\cite{JSAC_2022_An_Stacked}.
\end{itemize}

Once appropriately configured, SIMs can be utilized to perform a variety of inference tasks by directly processing information carried by electromagnetic waves. To elaborate on the unique capabilities of SIMs, we next provide a critical appraisal of its potential applications in communication (Sec. \ref{sec4}), sensing (Sec. \ref{sec5}), and computing (Sec. \ref{sec6}) systems.

\section{SIM for Wireless Communications}\label{sec4}
In wireless communication systems, modulation, filtering, and beamforming are particularly crucial for effectively conveying information, for shaping desired signal waveforms, and for directing beams toward intended users~\cite{BOOK_2005_Tse_Fundamentals}. As a promising alternative to traditional baseband processing techniques, SIMs can accomplish these tasks more efficiently in the wave domain.
\subsection{MIMO Precoding}
In the pioneering work~\cite{JSAC_2022_An_Stacked}, a pair of SIMs was employed at the transmitter and receiver for automatically implementing MIMO precoding and combining, as the electromagnetic waves propagate through the well-configured metasurfaces. With the use of SIMs, MIMO transceivers can eliminate the need for complex matrix multiplications in the digital domain. This substantially reduces processing latency, hardware costs, and energy usage compared to conventional MIMO systems~\cite{JSAC_2022_An_Stacked}. Furthermore, the hardware complexity can be reduced by transforming complex multi-stream detection schemes into parallel single-stream detection schemes. Building on this concept, the authors of~\cite{ICC_2023_An_Stacked, arXiv_2023_Jiancheng_Stacked} applied the SIM-based transceiver design concept to multiuser multiple-input single-output (MISO) downlink communications. In this scenario, a SIM was integrated with the radome of the base station (BS) to carry out transmit beamforming directly in the electromagnetic wave domain. To maximize the sum rate of all users, an optimization problem was formulated by jointly optimizing transmit power allocation and analog beamforming in the wave domain, subject to constraints on the transmit power budget and phase shifts. By decomposing the joint optimization problem into two sub-problems, the power allocation sub-problem was addressed using the popular iterative water-filling algorithm, while the SIM was configured through successive refinement for discrete phase shift tuning~\cite{arXiv_2023_Jiancheng_Stacked} and gradient ascent (GA) for continuous phase shift tuning~\cite{ICC_2023_An_Stacked}.

\begin{figure*}[!t]
\centering
\includegraphics[width=18cm]{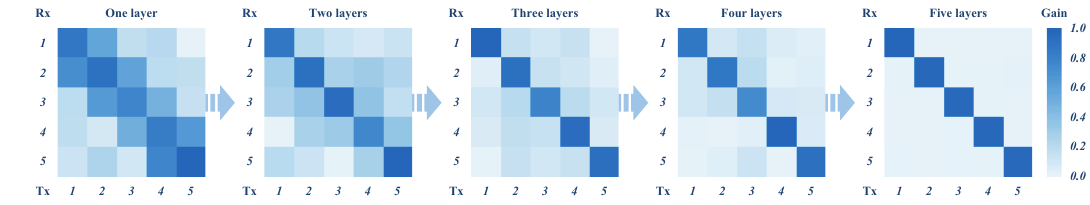}
\caption{Example of a $5 \times 5$ MIMO channel, where a pair of SIMs are employed and optimized to perform precoding and combining in the wave domain. Therefore, the five data streams can be transmitted and received in parallel through the corresponding antennas, without interfering with each other.}
\label{fig_7}
\end{figure*}
\begin{table*}[!t]
\centering
\renewcommand\arraystretch{1.25}
\caption{A comparison of SIM-based MIMO systems with their conventional counterparts in terms of computational complexity and hardware architecture.}
\label{tab_2}
\begin{threeparttable}
\begin{tabular}{>{\columncolor{red!15}}l|>{\columncolor{orange!15}}c|>{\columncolor{yellow!15}}c|>{\columncolor{green!15}}c|>{\columncolor{cyan!15}}c|>{\columncolor{blue!15}}l|>{\columncolor{violet!15}}c|>{\columncolor{purple!15}}l|>{\columncolor{darkgray!15}}l}
\hline
Scenario & $\#$ of sub-carriers & $\#$ of UEs & $\#$ of BSs & SIM & Reference & Year & $\#$ of RF chains & Complexity order \\ \hline\hline
 & & & & \faTimes & Goldsmith \emph{et al.}~\cite{JSAC_2003_Goldsmith_Capacity} & 2003 & $N_{\text{RF}} = N_{\text{Tx}}$ & $\mathcal{O}\left ( N_{\text{Tx}}N_{\text{Stream}} \right )$ \\ \cline{5-9}
\multirow{-2}{*}{MIMO} & \multirow{-2}{*}{$N_{\text{SC}} = 1$}& \multirow{-2}{*}{$N_{\text{UE}} = 1$}& \multirow{-2}{*}{$N_{\text{BS}} = 1$} & \faCheck & An \emph{et al.}~\cite{JSAC_2022_An_Stacked} & 2023& $N_{\text{RF}} = N_{\text{Stream}}$ & $\mathcal{O}\left ( 1 \right )$ \\ \hline\hline
 & & & & \faTimes&Spencer \emph{et al.}~\cite{TSP_2004_Spencer_Zero} & 2004 & $N_{\text{RF}} = N_{\text{Tx}}$ & $\mathcal{O}\left ( N_{\text{Tx}}N_{\text{UE}} \right )$\\ \cline{5-9}
 \multirow{-2}{*}{Multiuser} & \multirow{-2}{*}{$N_{\text{SC}} = 1$} & \multirow{-2}{*}{$N_{\text{UE}} > 1$}& \multirow{-2}{*}{$N_{\text{BS}} = 1$} & \faCheck&An \emph{et al.}~\cite{arXiv_2023_Jiancheng_Stacked} & 2024& $N_{\text{RF}} = N_{\text{UE}}$ & $\mathcal{O}\left ( 1 \right )$\\ \hline\hline
 & & & & \faTimes & Ngo \emph{et al.}~\cite{TWC_2017_Ngo_Cell} & 2017& $N_{\text{RF}} = N_{\text{BS}}N_{\text{Tx}}$ & $\mathcal{O}\left ( N_{\text{BS}}N_{\text{Tx}}N_{\text{UE}} \right )$ \\ \cline{5-9}
 \multirow{-2}{*}{Cell-free} & \multirow{-2}{*}{$N_{\text{SC}} = 1$} & \multirow{-2}{*}{$N_{\text{UE}} > 1$}& \multirow{-2}{*}{$N_{\text{BS}} > 1$} & \faCheck& Shi \emph{et al.}~\cite{arXiv_2024_Enyu_Harnessing} & 2024 & $N_{\text{RF}} \leq N_{\text{BS}}N_{\text{UE}}$ & $\mathcal{O}\left ( 1 \right )$ \\ \hline\hline
 & & & & \faTimes&Bolcskei \emph{et al.}~\cite{TCOM_2002_Bolcskei_On} & 2002& $N_{\text{RF}} = N_{\text{Tx}}$ & $\mathcal{O}\left ( N_{\text{Tx}}N_{\text{Stream}} + N_{\text{Tx}}\log N_{\text{SC}} \right )$ \\ \cline{5-9}
 \multirow{-2}{*}{OFDM} & \multirow{-2}{*}{$N_{\text{SC}} > 1$} & \multirow{-2}{*}{$N_{\text{UE}} = 1$}& \multirow{-2}{*}{$N_{\text{BS}} = 1$} & \faCheck&Li \emph{et al.}~\cite{TWC_2025_Li_Stacked} & 2025& $N_{\text{RF}} = N_{\text{Stream}}$ & $\mathcal{O}\left ( N_{\text{Stream}}\log N_{\text{SC}} \right )$ \\ \hline
\end{tabular}
\begin{tablenotes}
\item $\diamond$ $N_{\text{Tx}}$: Number of transmit antennas at each BS;\quad $\diamond$ $N_{\text{Stream}}$: Number of data streams.
\end{tablenotes}
\hrulefill
\end{threeparttable}\vspace{-0.4cm}
\end{table*}

To elucidate the underlying architectural paradigm shifts, Fig.~\ref{fig_6} contrasts a conventional MIMO framework with a SIM-based MIMO system. In conventional MIMO deployments, multiple data streams are spatially multiplexed and superimposed via digital precoding at the baseband prior to transmission \cite{BOOK_2005_Tse_Fundamentals}. However, this digital synthesis inherently yields high peak-to-average power ratio (PAPR) waveforms, which consequently necessitate high-resolution digital-to-analog converters (DACs) and analog-to-digital converters (ADCs) to mitigate quantization noise. Furthermore, severe PAPR scaling forces the transmitter's power amplifiers (PAs) and the receiver's low-noise amplifiers (LNAs) to operate with substantial back-off, drastically diminishing their power efficiency \cite{arXiv_2024_Jiancheng_Emerging}. In stark contrast, SIM-based MIMO systems shift the burden of transmit precoding and receive combining entirely into the electromagnetic wave domain, bypassing the requirement for dense digital processing blocks altogether \cite{JSAC_2022_An_Stacked}. Under this paradigm, each radio-frequency (RF) chain maps to a single antenna element transmitting an un-superimposed data stream. When coupled with low-order modulation schemes, the resulting constant-envelope or low-PAPR waveforms relax the resolution constraints on the underlying DACs and ADCs. Furthermore, the total RF chain count is compressed to match the number of independent data streams (or users), yielding profound savings in both transceiver hardware expenditure and operational energy dissipation \cite{WC_2024_An_Stacked, arXiv_2024_Hao_Stacked}.

While practical SIM architectures may introduce localized hardware overhead—such as the PIN diodes required to dynamically modulate the phase-shift profiles—this expenditure is heavily offset by the sweeping simplification of the transceiver front-end, especially when compared to massive MIMO arrays reliant on fully digital beamforming. Beyond hardware topology, a pivotal operational distinction arises at the execution layer. Specifically, traditional digital beamforming demands repetitive, power-intensive matrix multiplications and high-fidelity data conversions for every consecutive symbol transmission. Conversely, once the SIM's spatial phase profiles are configured, the entire precoding operation is executed passively and instantaneously via physical wave diffraction. Because the SIM requires no active runtime computational energy to route signals, its systemic and computational advantages over traditional digital MIMO scale monotonically with the volume of transmitted data.

Additionally, Fig.~\ref{fig_7} illustrates the actions of an end-to-end MIMO channel utilizing the SIM-based architecture of Fig.~\ref{fig_6}(b). A pair of SIMs is employed to mitigate interference among five data streams. The system operates at $28$ GHz. Moreover, each metasurface consists of $10 \times 10$ meta-atoms arranged with half-wavelength spacing. The axial spacing between adjacent metasurface layers is set to one wavelength, while all other simulation parameters are consistent with those outlined in~\cite{WC_2024_An_Stacked}. The transmission coefficients of the transmitting and receiving SIMs are optimized using the gradient descent (GD) method described in~\cite{JSAC_2022_An_Stacked}. As seen from Fig.~\ref{fig_7}, by utilizing an adequate number of metasurface layers, it becomes possible to create multiple interference-free channels, even in the absence of digital precoding and combining.

\begin{table*}[!t]
\centering
\scriptsize
\renewcommand\arraystretch{1.25}
\caption{A survey of existing studies on SIM, with a focus on optimization problem formulation and solver.}
\label{tab_3}
\begin{tabular}{c|c|l|l|l||l}
\hline
 Ref. & Link & Optimization problem & Solution & Characteristic & $\circledast$ Variables \& $\circledcirc$ Constraints \\ \hline\hline
\cite{arXiv_2025_Eduard_Scaling} & ----- & $\begin{aligned}&\min\ \sum_{s=1}^{S}\sum_{\tilde{s}\neq s}^{S}\left | h_{s,\tilde{s}}\right |^{2}\\ & \circledast \left\{ \text{v}_{1}\right\} \\ & \circledcirc \left\{ \text{C}_{1}\right\}\end{aligned}$ & \makecell[l]{Layer-by-layer optimization,\\closed-form solution} & Two SIMs & \multirow{16}{*}{$\begin{aligned}\star\quad &\text{SIM}\\\text{v}_{1} :&\ \psi _{n}^{l},\ \forall n,\ \forall l\\\text{v}_{2} :&\ \psi _{n}^{l}\left [ t \right ],\ \forall n,\ \forall l,\ \forall t\\\text{v}_{3} :&\ \varsigma _{n}^{l},\ \forall n,\ \forall l\\ \\\ast\quad &\text{Transmitter}\\\text{v}_{4} :&\ p_{k},\ \forall k\\\text{v}_{5} :&\ p_{k}\left [ t \right ],\ \forall k,\ \forall t\\\text{v}_{6} :&\ \mathbf{Q}\\ \\\circ\quad &\text{RIS}\\ \text{v}_{7}: &\ \theta_{m},\ \forall m\\\text{v}_{8}: &\ \varrho_{m},\ \forall m \\ \\\diamond\quad &\text{UAV}\\ \text{v}_{9} :&\ \mathbf{q}_{k}\left [ t \right ],\ \forall k,\ \forall t\\ \\ \\\star\quad &\text{SIM}\\\text{C}_{1}: &\ \psi _{n}^{l}\in \left [ 0,2\pi \right ),\ \forall n,\ \forall l\\ \text{C}_{2}: &\ \psi _{n}^{l}\left [ t \right ]\in \left [ 0,2\pi \right ),\ \forall n,\ \forall l,\ \forall t \\\text{C}_{3} :&\ \varsigma _{n}^{l}\leq \varsigma _{\max},\ \forall n,\ \forall l \\ \\ \ast\quad &\text{Transmitter}\\\text{C}_{4} :&\ \sum\limits_{s=1}^{S}p_{s}\leq P_{\max}\\\text{C}_{5} :&\ \sum\limits_{k=1}^{K}p_{k}\left [ t \right ]\leq P_{\max},\ \forall t\\\text{C}_{6} :&\ p_{s}\geq 0,\ \forall s\\\text{C}_{7} :&\ p_{k}\left [ t \right ]\geq 0,\ \forall k,\ \forall t\\\text{C}_{8} :&\ \text{tr}\left ( \mathbf{Q} \right )\leq P_{\max} \\ \text{C}_{9} :&\ \mathbf{Q} \succeq \mathbf{0}\\ \\\circ\quad &\text{RIS}\\\text{C}_{10} :&\ \theta _{m}\in \left [ 0,2\pi \right ),\ \forall m\\\text{C}_{11}:&\ \varrho _{m}\leq \varrho _{\text{max}},\ \forall m \\ \\ \diamond\quad &\text{UAV}\\\text{C}_{12} :&\ \left\| \mathbf{q}_{k}\left [ t+1 \right ]- \mathbf{q}_k\left [ t \right ] \right\|\leq \upsilon _{\max}\tau,\ \forall k,\ \forall t \\ \text{C}_{13} :&\ \mathbf{q}_k\left [ t \right ]\in \mathcal{Q},\ \forall k,\ \forall t \\ \\\bullet\quad &\text{QoS / QoE}\\\text{C}_{14}:&\ R_{k} \geq R_{\text{min}},\ \forall k\\\text{C}_{15} :&\ R_{k}\left [ t \right ] \geq R_{\text{min}},\ \forall t\\\text{C}_{16} :&\ \text{O}_{k}\geq \text{O}_{\min},\ \forall k \end{aligned}$}\\ \cline{1-5}
\cite{TWC_2024_Papazafeiropoulos_Achievable} & ----- & $\begin{aligned}&\max\ C\\& \circledast \left\{ \text{v}_{1},\text{v}_{6}\right\}\\& \circledcirc \left\{ \text{C}_{1},\text{C}_{8},\text{C}_{9}\right\}\end{aligned}$ & Projected GA & Two SIMs & \\ \cline{1-5}
\cite{CL_2024_Stefan_Mutual} & ----- & $\begin{aligned}&\max\ R_{\text{C}}\\& \circledast \left\{ \text{v}_{1},\text{v}_{6}\right\}\\& \circledcirc \left\{ \text{C}_{1},\text{C}_{8},\text{C}_{9}\right\}\end{aligned}$ & Projected GA & \makecell[l]{Two SIMs,\\discrete alphabet} & \\ \cline{1-5}
\cite{arXiv_2024_Hao_Multiuser} & DL & \multirow{3}{*}{\makecell[l]{\\$\begin{aligned}&\max\ \sum\limits_{k=1}^{K}R_{k}\\& \circledast \left\{ \text{v}_{1},\text{v}_{4}\right\}\\& \circledcirc \left\{ \text{C}_{1},\text{C}_{4},\text{C}_{6}\right\}\end{aligned}$}} & \makecell[l]{\\DDPG} & \makecell[l]{\\\emph{n.a.}} & \\ \cline{1-2}\cline{4-5}
\cite{arXiv_2024_Xiaolei_Joint} & DL & & \makecell[l]{\\TD3} & \makecell[l]{\\\emph{n.a.}} & \\ \cline{1-2}\cline{4-5}
\cite{WCL_2024_Papazafeiropoulos_Achievable} & DL & & \makecell[l]{\\$\underline{\text{AO}}\left\{\begin{aligned}
 \text{v}_{1}&\text{: Projected GA}\\\text{v}_{4}&\text{: Weighted MMSE}
\end{aligned}\right.$} & \makecell[l]{Statistical CSI} & \\ \cline{1-5}
\cite{arXiv_2024_Donatella_Design} & DL & $\begin{aligned}&\max\ \sum\limits_{k=1}^{K}R_{k}\\& \circledast \left\{ \text{v}_{1},\text{v}_{3},\text{v}_{4}\right\}\\& \circledcirc \left\{ \text{C}_{1},\text{C}_{3},\text{C}_{4},\text{C}_{6}\right\}\end{aligned}$ & $\underline{\text{AO}}\left\{\begin{aligned}
 \text{v}_{1},\text{v}_{3}&\text{: Projected GA}\\\text{v}_{4}&\text{: Iterative WF}
\end{aligned}\right.$ & \makecell[l]{Active layers,\\user scheduling} & \\ \cline{1-5}
\cite{arXiv_2024_Mohammadzadeh_Meta} & DL & \multirow{2}{*}{$\begin{aligned}&\max\ \sum\limits_{k=1}^{K}R_{k}\\& \circledast \left\{ \text{v}_{1},\text{v}_{4}, \text{v}_{7}\right\}\\& \circledcirc \left\{ \text{C}_{1},\text{C}_{4},\text{C}_{6},\text{C}_{10},\text{C}_{14}\right\}\end{aligned}$} & \makecell[l]{\\TD3, meta-learning\\ \\} & \makecell[l]{\\Passive RIS,\\user mobility\\ \\} & \\ \cline{1-2}\cline{4-5}
\cite{arXiv_2024_Zarini_Interplay} & DL & & \makecell[l]{\\NAC, meta-learning\\ \\} & \makecell[l]{\\STAR-RIS,\\user mobility\\ \\}\\ \cline{1-5}
\cite{TCOM_2025_Fang_Stacked} & DL & $\begin{aligned}&\max\ \left ( \prod \limits_{k=1}^{K}R_{k} \right )^{1/K}\\& \circledast \left\{ \text{v}_{1},\text{v}_{4}\right\}\\& \circledcirc \left\{ \text{C}_{1},\text{C}_{4},\text{C}_{6}\right\}\end{aligned}$ & Consensus ADMM & \makecell[l]{Rate fairness} & \\ \cline{1-5}
\cite{arXiv_2024_Zarini_QoE} & DL & $\begin{aligned}&\max\ \sum\limits_{k=1}^{K}\text{O}_{k}\\& \circledast \left\{ \text{v}_{1},\text{v}_{4}\right\}\\& \circledcirc \left\{ \text{C}_{1},\text{C}_{4},\text{C}_{6},\text{C}_{16}\right\}\end{aligned}$ & CQL, meta-learning & \makecell[l]{QoE,\\user mobility} & \\ \cline{1-5}
\cite{arXiv_2024_Zarini_On} & DL & $\begin{aligned}&\max\ N_{\text{UE}}\\& \circledast \left\{ \text{v}_{1},\text{v}_{4},\text{v}_{8}\right\}\\& \circledcirc \left\{ \text{C}_{1},\text{C}_{4},\text{C}_{6},\text{C}_{10},\text{C}_{11},\text{C}_{14}\right\}\end{aligned}$ & \makecell[l]{MPO, meta-learning} & \makecell[l]{Active RIS,\\user mobility} & \\\cline{1-5}
\cite{arXiv_2024_Anastasios_Performance} & UL & $\begin{aligned}&\max\ \sum\limits_{k=1}^{K}R_{k}\\& \circledast \left\{ \text{v}_{1}\right\}\\& \circledcirc \left\{ \text{C}_{1}\right\}\end{aligned}$ & Projected GA & Ambient SIM & \\ \cline{1-5}
\cite{arXiv_2024_On_Zarini} & DL & \multirow{2}{*}{$\begin{aligned}&\max\ \sum\limits_{t=1}^{T}\sum\limits_{k=1}^{K}R_{k}\left [ t \right ]\\& \circledast \left\{ \text{v}_{2},\text{v}_{5},\text{v}_{9}\right\}\\& \circledcirc \left\{ \text{C}_{2},\text{C}_{5},\text{C}_{7},\text{C}_{12},\text{C}_{13},\text{C}_{15}\right\}\end{aligned}$} & \makecell[l]{\\D4PG, meta-learning\\} & \makecell[l]{\\UAV,\\user mobility} &\\ \cline{1-2}\cline{4-5}
\cite{TNSE_2026_Xiong_Digital} & DL & & $\underline{\text{AO}}\left\{\begin{aligned}
 &\underline{\text{AO}}\left\{\begin{aligned}
 &\text{v}_2\text{: GA}\\&\text{v}_5\text{: FP}
\end{aligned}\right.\\&\text{v}_9\text{: DQN}
\end{aligned}\right.$ & \makecell[l]{Digital twin,\\flight control} & \\ \cline{1-5}
\cite{TIFS_2024_Niu_On} & DL & $\begin{aligned}&\min\ \left\| \mathbf{x} - \hat{\mathbf{x}}\right\|^{2}\\& \circledast \left\{ \text{v}_{1}\right\}\\& \circledcirc \left\{ \text{C}_{1}\right\}\end{aligned}$ & $\begin{aligned}&\textrm{Layer-by-layer optimization},\\&\textrm{closed-form solution}\end{aligned}$ & \makecell[l]{PLS,\\reduced power loss} & \\ \hline
\end{tabular}\vspace{-0.4cm}
\end{table*}

\begin{table*}[!t]
\centering
\caption{The major symbols in Table \ref{tab_3} and their meanings.}
\label{tab_4}
\renewcommand\arraystretch{1.25}
\begin{tabular}{l|l||l|l}
\hline
Symbol & Meaning & Symbol & Meaning \\ \hline\hline
$p_{k}\left [ t \right ]$ & Transmit power allocated to the $k$-th user at time slot $t$ & $\upsilon _{\max}$ & Maximum flight velocity\\ \hline
$\psi _{n}^{l}\left [ t \right ]$ & Phase shift of the $n$-th meta-atom on the $l$-th metasurface layer at time slot $t$ & $\tau$ & Duration of each time slot \\ \hline
$\mathbf{q}_{k}\left [ t \right ]$ & Position of the $k$-th user at time slot $t$ & $\mathcal{Q}$ & Air corridor \\ \hline
$R_{k}\left [ t \right ]$ & Achievable rate of the $k$-th user at time slot $t$ & $T$ & Number of time slots \\ \hline
$\varrho_{m}$ & Amplitude of the $m$-th RIS element & $\varrho _{\text{max}}$ & Maximum amplitude provided by the active load \\ \hline
$\varsigma _{n}^{l}$ & Amplitude response of the $n$-th meta-atom on the $l$-th metasurface layer & $\varsigma _{\max}$& Maximum amplitude response \\ \hline
$p_k$ & Power allocated to the $k$-th user & $P_{\max}$ & Maximum transmit power \\ \hline
$\text{O}_{k}$ & Mean opinion score of the web service for user $k$ & $\text{O}_{\min}$& Minimum QoE level \\ \hline
$R_{k}$ & Achievable rate of the $k$-th user & $R_{\text{min}}$ & QoS requirement \\ \hline
$p_s$ & Power allocated to the $s$-th data stream & $K$ & Number of users\\ \hline
$\psi _{n}^{l}$ & Phase shift of the $n$-th meta-atom on the $l$-th metasurface layer & $\theta_{m}$& Phase shift of the $m$-th RIS element \\ \hline
$\mathbf{Q}$ & Transmit covariance matrix & $N_{\text{UE}}$ & Number of served users \\ \hline
$h_{s,\tilde{s}}$ & End-to-end channel from the $\tilde{s}$-th transmit antenna to the $s$-th receive antenna & $S$ & Number of data streams \\ \hline
$\mathbf{x}$ & The actual signal output by the last layer of the SIM & $R_{\text{C}}$ & Channel cutoff rate \\ \hline
$\hat{\mathbf{x}}$ & The desired signal output by the last layer of the SIM & $C$ & Channel capacity\\ \hline
\end{tabular}\vspace{-0.4cm}
\end{table*}

Over the past few years, the concept of utilizing SIMs to implement precoding in the wave domain has gained considerable attention and has been applied to various communication systems \cite{WCL_2026_Bahingayi_A}. Next, we will review the notable progress in this area from four key perspectives.
\begin{itemize}
\item \textbf{From Far-Field Communications to Near-Field Communications:} In order to simultaneously achieve high data throughput and massive connectivity in next-generation wireless networks, extremely large-scale antenna arrays (ELAAs) and metasurfaces are expected to be widely deployed~\cite{WC_2024_An_Near}. As a consequence, wireless communications will predominantly operate in the radiating near-field region. Although the fundamental system design and signal processing techniques for near-field and far-field communications are similar, the unique characteristics of spherical wavefronts in near-field propagation offer enhanced spatial multiplexing gain~\cite{WC_2024_An_Near, TCOM_2025_Li_Stacked}. Furthermore, in addition to angular selectivity, they also allow `spot-light-like' beam-focusing to a specific location, hence allowing for the distinction of users in the distance domain. However, this advantage necessitates complex matrix operations for baseband beamfocusing at the transmitter, which can lead to high computational complexity and processing delays when using ELAAs or metasurfaces. In this context, applying a SIM to perform near-field beamfocusing in the wave domain becomes increasingly appealing. Building on this idea, Jia \emph{et al.}~\cite{VTC_2024_Jia_Stacked} explored the potential of SIMs in multiuser MISO near-field communications, where a SIM was integrated into the radome of a BS to perform beamfocusing through wave manipulation. They employed a GD algorithm to configure the transmission coefficients of a SIM and customize an end-to-end channel that minimizes interference among users. Moreover, the authors of~\cite{WCL_2024_Papazafeiropoulos_Near} formulated an optimization problem aimed at maximizing the weighted sum rate of multiple near-field users by jointly designing the transmit power allocation and the phase shifts of the SIM. A block coordinate descent (BCD) algorithm was then applied to solve this problem.
\item \textbf{From Narrowband Beamforming to Wideband Beamforming:} Traditional wideband MIMO communication systems handle individual frequency bands independently through digital beamforming~\cite{TWC_2017_Park_Dynamic}. In contrast, a SIM must process multiple frequency sub-bands simultaneously using a single physical device, which can give rise to the detrimental beam squint phenomenon~\cite{TVT_2025_Ming_Flexible} — whereby the beam direction varies as a function of frequency. Nevertheless, unlike single-layer metasurfaces that exhibit a frequency-flat response, SIMs are inherently capable of producing frequency-selective transfer functions, owing to the multiple internal propagation paths established across their constituent layers. Inspired by this characteristic, Li \emph{et al.}~\cite{TWC_2025_Li_Stacked} explored the application of SIMs for fully-analog beamforming in MIMO-aided orthogonal frequency division multiplexing (OFDM) communication systems, which aims to achieve interference-free transmission across a broad frequency range. To achieve this, they employed a GD algorithm to determine the optimal SIM configuration and examined the relationship between the maximum effective transmission bandwidth and various physical parameters of the SIM, such as the number of metasurface layers and the metasurface aperture. Their numerical results demonstrated that optimizing the SIM for multiple frequency sub-carriers -- rather than only the center frequency -- significantly improved the channel capacity of SIM-aided MIMO-OFDM communication systems. Furthermore, in~\cite{arXiv_2025_Li_Stacked}, the same authors adopted index modulation (IM)-aided OFDM as the transmission waveform to illuminate SIM. In OFDM-IM, information is conveyed by selectively activating a subset of available subcarriers, with the specific indices of these active subcarriers implicitly conveying additional bits. This transmission strategy offers two key advantages: firstly, the selective subcarrier activation inherently reduces the PAPR of the broadband transmission waveform; secondly, the sparse subcarrier multiplexing among multiple users effectively expands the operational bandwidth over which a SIM can function effectively. To assess system reliability, the worst-case bit error rate (BER) was adopted as the performance metric, and a deep unfolding framework was developed for configuring the SIM parameters, while adaptively adjusting the step size at each iteration.
\item \textbf{From Single BS to Multiple BSs:} The capacity of wireless networks can be enhanced by deploying more BSs or access points (APs); however, this approach also increases both the hardware costs and energy consumption accordingly~\cite{Proc_2022_Shi_RIS}. To tackle this challenge, recent studies have investigated the integration of low-cost SIMs with APs in cell-free massive MIMO systems. In downlink scenarios, Hu \emph{et al.}~\cite{TVT_2024_Hu_Joint} developed a method that jointly optimizes AP power allocation and SIM-based wave-domain beamforming to maximize the sum rate. For uplink applications, Shi \emph{et al.}~\cite{APwCS_2024_Shi_Uplink, arXiv_2024_Enyu_Harnessing} developed a two-stage transmission framework. The first stage employs a greedy algorithm to minimize interference among pilot signals during channel estimation. In the second stage, the SIM is configured to achieve wave-domain beamforming, and a power allocation algorithm is developed for maximizing the worst-case spectral efficiency of user equipment (UE). Additionally, Li \emph{et al.}~\cite{TCOM_2024_Li_Stacked} optimized the SIM coefficients for accurately recovering the symbols for each UE. The central processing unit (CPU) then combined the local detection results using weights designed based on the minimum mean square error (MMSE) criteria, taking into account the impact of hardware impairments. These studies demonstrate that increasing the number of metasurface layers can significantly reduce inter-user interference by effectively leveraging wave manipulation. More recently, Park \emph{et al.}~\cite{TWC_2025_Park_SIM} proposed a hybrid beamforming approach that combines wave-domain beamforming through SIMs with standard digital processing. Specifically, an optimization framework was developed for jointly configuring the digital beamforming, wave-domain beamforming, and fronthaul compression for maximizing the weighted sum-rate of both the uplink and downlink, when fronthaul capacity is limited.
\begin{figure*}[!t]
\centering
\includegraphics[width=18cm]{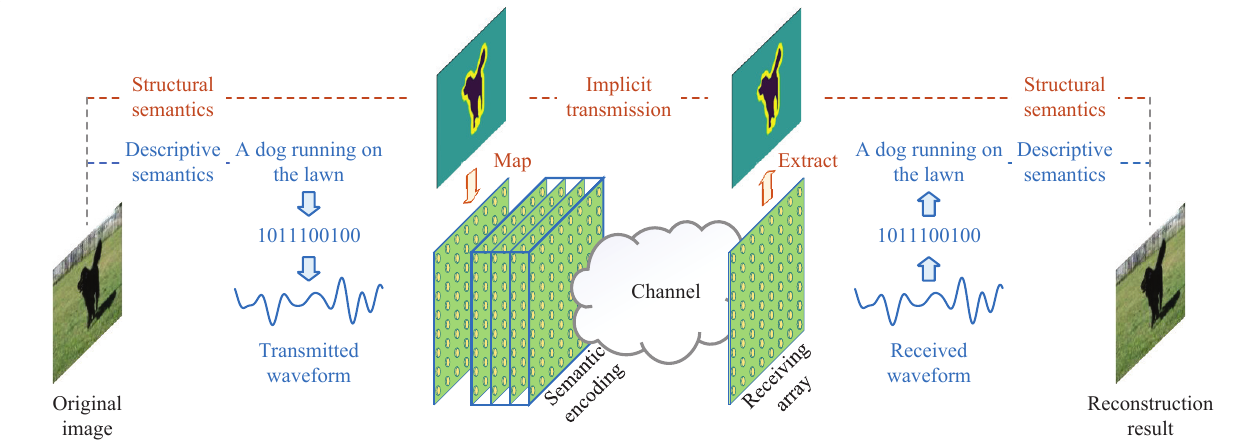}
\caption{An application of SIMs in bimodal semantic communication systems, where the structural semantic information is implicitly conveyed by imaging the edge patterns onto the receiving array, while the descriptive semantic information is transmitted using traditional phase-amplitude modulation techniques. The receiver subsequently reconstructs the original image by synthesizing the received bimodal semantic information.}
\label{fig_8}\vspace{-0.4cm}
\end{figure*}
\item \textbf{From Classical Configuration to Intelligent Configuration:} Achieving the targeted multi-user MIMO multiplexing gains in the wave domain necessitates precise SIM configuration. However, the high-dimensional parameter space inherent in multi-layer SIM architectures introduces prohibitive computational complexity and processing latency for conventional optimization methods. To circumvent these bottlenecks, recent research paradigms have pivoted from classical configuration techniques toward intelligent, data-driven SIM control strategies. For instance, Liu \emph{et al.} \cite{arXiv_2024_Hao_Multiuser} proposed a customized model-free DRL approach based on the deep deterministic policy gradient (DDPG) framework to maximize the downlink sum rate through the joint optimization of continuous SIM phase shifts and BS transmit power allocation. By integrating an attenuated whitening process to enhance exploration, this framework bypasses the need for extensive offline datasets while maintaining low online computational complexity. Extending intelligent control to cell-free massive MIMO deployments, Zhu \emph{et al.} \cite{arXiv_2025_Zhu_Joint} addressed the dimensionality challenges of joint AP power allocation and SIM configuration by introducing a distributed multi-agent reinforcement learning (MARL) framework. Operating under a centralized training with decentralized execution protocol, their design leverages a noisy value method paired with a recurrent policy network to maximize the system's sum spectral efficiency. Furthermore, to tackle the complex, non-convex joint precoding and phase-shift optimization problem for physical layer security (PLS) in multi-user networks, Shi \emph{et al.} \cite{arXiv_2026_Shi_Low} developed an innovative manifold-enhanced heterogeneous multi-agent continual learning (MHACL) framework. By modeling each SIM layer as an autonomous agent, MHACL counters environmental non-stationarity and distribution shifts through dynamic gradient masking, prioritized experience replay, and geometric regularization—effectively consolidating historical knowledge without catastrophic forgetting. To further alleviate the computational burden, a low-complexity learning template was also derived that embeds the phase configuration space directly into a structured product manifold.
\end{itemize}

To further illustrate the benefits of SIM-based MIMO systems, Table \ref{tab_2} compares the number of RF chains required and the computational complexity involved in implementing beamforming in these communication systems versus their conventional counterparts. As shown in Table \ref{tab_2}, the utilization of SIM not only facilitates efficient beamforming in the wave domain but also significantly reduces both the hardware costs and computation demands. Furthermore, recent literature has extensively explored the integration of SIMs with other vanguard wireless technologies, such as DRL~\cite{arXiv_2024_Hao_Multiuser, arXiv_2024_Xiaolei_Joint, arXiv_2024_Mohammadzadeh_Meta, ICC_2024_Liu_DRL, ATC_2025_Hoang_Secure, TVT_2025_Yang_Low}, reconfigurable intelligent surfaces (RIS)~\cite{arXiv_2024_Zarini_On, arXiv_2024_Zarini_Interplay, arXiv_2024_Anastasios_Performance}, rate splitting multiple access (RSMA)~\cite{CL_2025_Liu_Sum, JSAC_2025_Sun_Dual}, and UAV~\cite{arXiv_2024_On_Zarini, TNSE_2026_Xiong_Digital}. Other sophisticated applications include PLS~\cite{APWCS_2024_Niu_Enhancing, TIFS_2024_Niu_On}, non-orthogonal multiple access (NOMA)~\cite{TAES_2025_Yu_Energy}, and simultaneous wireless information and power transfer (SWIPT)~\cite{arXiv_2024_Amiri_Stacked}, to fully exploit their wave-domain signal processing capabilities.

To provide a comprehensive understanding of these recent advances from a mathematical perspective, Table \ref{tab_3} systematically summarizes typical optimization problem formulations, explicitly delineating the prevailing optimization objectives, key optimization variables, operational constraints, as well as common solvers. Additionally, the relevant mathematical symbols and their definitions are listed in Table \ref{tab_4}. Observing from Table \ref{tab_3} offers key insights across several dimensions to spark future research avenues. \emph{First}, regarding optimization formulations, prevailing models predominantly focus on maximizing the achievable rate or utilizing the electromagnetic response of the SIM to synthesize a target diagonal channel matrix. Conversely, direct evaluation of system-level performance metrics, such as the BER, remains remarkably scarce. Furthermore, multi-user fairness—particularly the capacity of the SIM to mitigate co-channel interference within near-far user topologies—warrants more profound investigation. \emph{Second}, given that the SIM represents a novel physical computing paradigm, its operational efficacy is inherently coupled with hardware precision. Consequently, evaluating SIM performance under realistic hardware impairments, including discrete phase-shift quantization, structural inter-layer misalignment, and mutual coupling, is a crucial open challenge. \emph{Third}, existing algorithmic solutions are largely a legacy continuation of conventional RIS optimization techniques. Exploiting the innate, wave-based online analog computing capabilities of the SIM, in conjunction with limited feedback mechanisms, could drastically accelerate training and configuration processes, thereby unlocking its full potential in highly dynamic deployment scenarios.

\subsection{Semantic Encoding}
Semantic communication is a promising paradigm that can significantly enhance communication efficiency. In task-oriented semantic communication systems, a SIM can be utilized as a semantic encoder. For example, the authors of~\cite{Arxiv_2024_Huang_Stacked} demonstrated this concept by investigating a classification task in which a SIM was strategically positioned before the transmit antenna. The input layer of the SIM was utilized for source encoding, while the remaining layers formed a DNN for semantic encoding, which aims for transforming the signals traversing the input layer into a unique beam directed toward the receiving antenna corresponding to a specific class. In contrast to conventional communication systems that transmit modulated signals containing either encoded information or compressed semantic information, the transmit antenna of SIM-based semantic communication systems simply emits unmodulated electromagnetic carrier waves~\cite{Arxiv_2024_Huang_Stacked}. At the receiver, the image class is recognized by probing the signal magnitude distributions across the receiving array, which substantially reduces the required signal processing complexity. Notably, both the source and semantic encoding occur automatically as the electromagnetic waves propagate through the stratified structure. By adopting a similar philosophy, Liu \emph{et al.}~\cite{ICC_2025_Liu_A} developed a disaster recognition method that consists of a SIM mounted on a drone to process semantic information encoded in electromagnetic waves, while dissipating low energy and at extremely fast speeds. At the ground station, the electronic network further processes the received signals to improve the system's ability to make more accurate disaster predictions.

When transmitting complex visual scenes, existing semantic communication systems encounter challenges related to substantial data transmission demands. To address this issue, Huang \emph{et al.}~\cite{WCL_2025_Huang_Stacked} leveraged a SIM placed in front of the transmit antenna to convey visual edge information from intricate scenes by directly mapping them using imaging techniques onto the receiver array, as illustrated in Fig.~\ref{fig_8}. Complementary textual scene descriptions are transmitted using traditional amplitude and phase modulation techniques. To realize the desired function, a customized mini-batch GD algorithm was developed for fine-tuning the transmission coefficients of meta-atoms within the SIM, thereby minimizing discrepancies between the received and desired visual patterns. Subsequently, a conditional generative adversarial network (GAN) processes the received visual patterns and text descriptions to reconstruct the complex original scene.

\section{SIM for Sensing}\label{sec5}
Wireless sensing is essential for acquiring critical information about targets of interest. Typical sensing applications encompass direction-of-arrival (DOA) estimation, object recognition, as well as target detection and tracking. In contrast to traditional sensing algorithms that rely on specific models and cannot meet real-time processing requirements, SIM provides a universal framework for extracting information by directly processing radio waves.

\begin{figure}[!t]
\centering
\includegraphics[width=8.5cm]{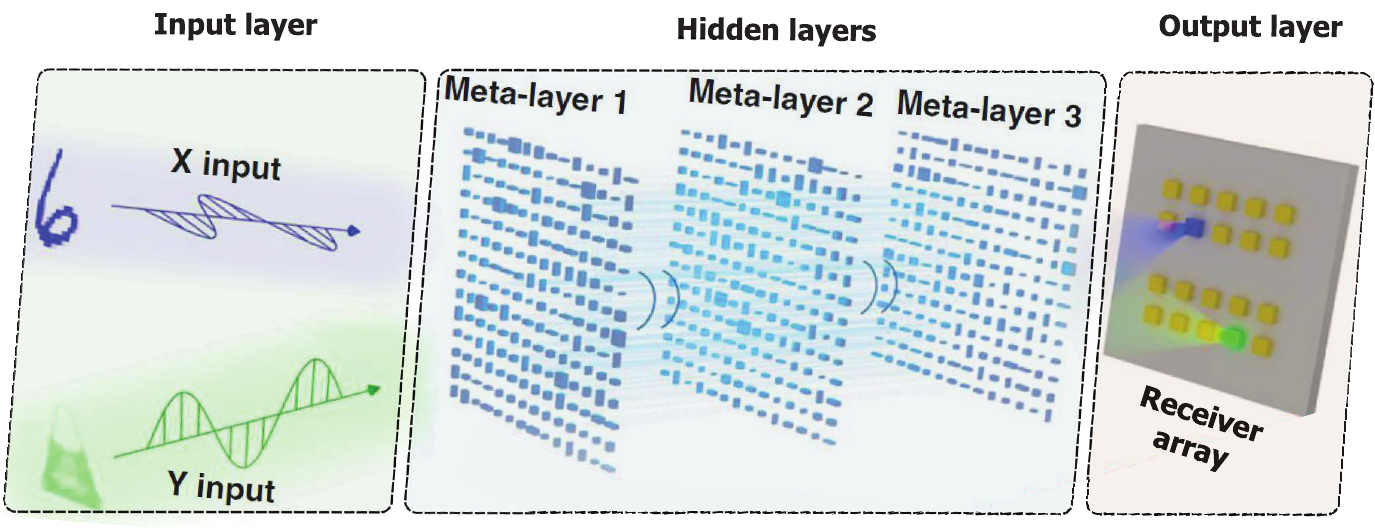}
\caption{A schematic diagram depicting the application of SIM to target recognition, with SIM operating as a physical neural network \cite{LSA_2022_Luo_Metasurface}.}
\label{fig_9}\vspace{-0.4cm}
\end{figure}
\subsection{Object Recognition}
Substantial research efforts have established that SIM has great potential in executing object recognition tasks~\cite{Sci_2018_Lin_All, NE_2022_Liu_A}. Building upon this foundation, Luo \emph{et al.}~\cite{LSA_2022_Luo_Metasurface} fabricated an advanced polarization-multiplexed SIM that operates in the visible spectrum, which demonstrates the capability to simultaneously perform dual recognition tasks: identifying handwritten digits and classifying fashionable items by using distinct polarization channels. As shown in Fig.~\ref{fig_9}, this innovative dual-channel object classification framework consists of three primary components:
\begin{itemize}
 \item \emph{Input layer:} A plane wave with appropriately engineered phase and amplitude distribution is used for encoding the object information across multiple channels.
 \item \emph{Hidden layers:} Each hidden layer contains asymmetric meta-atoms with tunable birefringence characteristics.
 \item \emph{Output layer:} The physical output plane is partitioned into several discrete detection regions, each corresponding to a particular object category.
\end{itemize}
By employing an error backpropagation approach, the multi-dimensional phase distributions are progressively updated, and, eventually, much of the energy of the input electromagnetic waves gleaned from different channels is correctly routed to their corresponding detection regions. The region having the highest intensity at the output layer indicates the class identified. Both computational simulations and experimental validation have confirmed that using a SIM with a deeper architecture can substantially enhance the recognition performance~\cite{LSA_2022_Luo_Metasurface}. For instance, the accuracy of recognizing $10$ handwritten digits improves from $65\%$ to $92\%$, when increasing the number of metasurface layers from one to seven (as illustrated in Fig. 2(a) of~\cite{LSA_2022_Luo_Metasurface}).

\begin{figure*}[!t]
\centering
\includegraphics[width=18 cm]{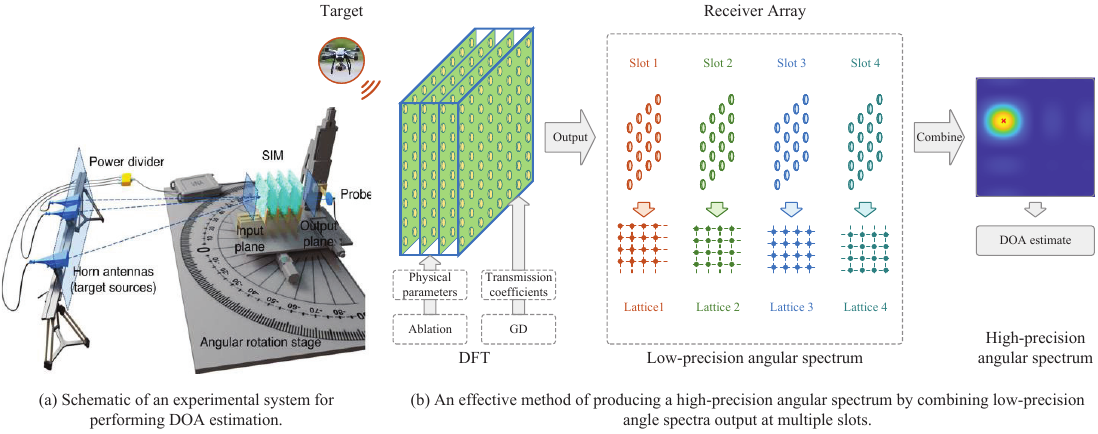}
\caption{Illustration of employing SIM for DOA estimation.}
\label{fig_11}\vspace{-0.4cm}
\end{figure*}
In addition to polarization multiplexing, a broadband SIM was developed in~\cite{SA_2021_Li_Spectrally} to perform statistical inference by encoding the spatial features of objects into spectral power distributions at preselected wavelengths, with each representing a different classification category. The SIM is trained to maximize the power of the specific spectral component corresponding to the true label of each object. As a result, a single-pixel detector at the output plane becomes sufficient for analyzing the power distribution across the spectrum and for performing object classification. Additionally, a compact fully-connected ANN having merely two hidden layers was trained for successfully reconstructing the image of the input objects based on the highly compressed spectral power representations~\cite{SA_2021_Li_Spectrally}, even if they were misclassified by the SIM. Recently, Hua \emph{et al.}~\cite{arXiv_2025_Hua_Implementing} presented AirFC, a representative paradigm that leverages SIM to enable over-the-air computation for low-latency and energy-efficient analog inference with enhanced image classification accuracy.

\subsection{Parameter Estimation}
Wireless sensing relying on radio wave propagation constitutes a fundamental enabler for diverse applications spanning communication, radar, and navigation systems~\cite{NP_2024_Huang_A, ICASSP_2025_Lin_UAV}. Conventional paradigms for target detection and parameter estimation rely upon an intricate process that involves the RF front-end circuitry to demodulate and sample multi-channel baseband signals, followed by computationally intensive algorithmic operations that inherently impose stringent limitations on sensing latency and energy efficiency (EE)~\cite{BOOK_2013_Poor_An}. By contrast, a SIM offers a compelling alternative by directly processing electromagnetic waves to estimate target parameters, thereby bypassing the conventional requirements for RF circuitry, ADCs, and digital signal processing infrastructure.

Recent years have witnessed growing research interest in leveraging SIMs for DOA estimation~\cite{LSA_2024_Gao_Super, LPR_2023_Huang_Diffraction,ICC_2024_An_Stacked, OJCOMS_2025_Javed_SIM}. The fundamental mechanism underlying SIM-based DOA estimation lies in spatially mapping the electromagnetic field distributions impinging from different source directions into distinct angular bins. Typically, a SIM is positioned in front of a detector array, where each detection region corresponds to a specific angular interval and captures the intensity of the received electromagnetic field. Through optimization of the control voltages applied to individual meta-atoms, the SIM can be trained to focus the energy of the incident waves arriving from targets at particular angles onto their corresponding detection regions at the output plane. In~\cite{LPR_2023_Huang_Diffraction}, Huang \emph{et al.} demonstrated the feasibility of SIM-based DOA estimation for multiple incident sources across wide frequency and angular ranges. By optimizing the physical parameters of the SIM, their sensing architecture creates distinct virtual channels that isolate and process different incoming wavefronts, enabling automatic beam classification and spatial routing based on incidence angle and frequency. Their experimental implementation employed a dual-layer SIM to detect multiple sources within the $8\sim12$ GHz frequency band for incident angles ranging from $-30^{\circ}$ to $30^{\circ}$. The results demonstrated significant reductions in both processing time and energy consumption compared to traditional DOA estimation techniques.

 \begin{figure}[!t]
\centering
\includegraphics[width=8cm]{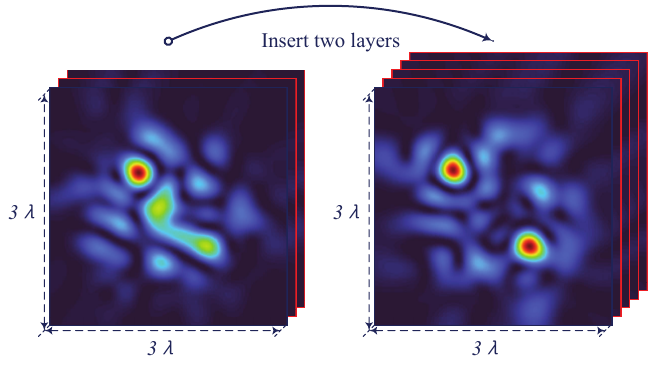}
\caption{Spatial frequency spectra produced by a two-layer and four-layer SIM, when two targets are present. The metasurface layers are spaced by one wavelength, each consisting of $10 \times 10$ half-wavelength spaced meta-atoms.}
\label{fig_10}\vspace{-0.4cm}
\end{figure}

\begin{table*}[!t]
\centering
\renewcommand\arraystretch{1.25}
\caption{The experimental energy distribution matrix and confusion matrix are evaluated on the single-target testing data set, where target angle interval $i$ represents the angular range of $\left [ \left ( i-6 \right )^{\circ},\left ( i-5 \right )^{\circ} \right )$.}
\begin{tabular}{l||cccccccccc||cccccccccc}
\hline
\multirow{2}{*}{} & \multicolumn{10}{c||}{\begin{tabular}[c]{@{}c@{}}\textbf{Energy distribution matrix}\\Angle interval\end{tabular}} & \multicolumn{10}{c}{\begin{tabular}[c]{@{}c@{}}\textbf{Confusion matrix}\\Angle interval\end{tabular}} \\ \cline{2-21} \cline{2-21}
 & \multicolumn{1}{c|}{1} & \multicolumn{1}{c|}{2} & \multicolumn{1}{c|}{3} & \multicolumn{1}{c|}{4} & \multicolumn{1}{c|}{5} & \multicolumn{1}{c|}{6} & \multicolumn{1}{c|}{7} & \multicolumn{1}{c|}{8} & \multicolumn{1}{c|}{9} & 10 & \multicolumn{1}{c|}{1} & \multicolumn{1}{c|}{2} & \multicolumn{1}{c|}{3} & \multicolumn{1}{c|}{4} & \multicolumn{1}{c|}{5} & \multicolumn{1}{c|}{6} & \multicolumn{1}{c|}{7} & \multicolumn{1}{c|}{8} & \multicolumn{1}{c|}{9} & 10 \\ \hline
Probe 1 & \multicolumn{1}{c|}{\textbf{0.44}} & \multicolumn{1}{c|}{0.19} & \multicolumn{1}{c|}{0.05} & \multicolumn{1}{c|}{0.01} & \multicolumn{1}{c|}{0.03} & \multicolumn{1}{c|}{0.03} & \multicolumn{1}{c|}{0.02} & \multicolumn{1}{c|}{0.01} & \multicolumn{1}{c|}{0.00} & 0.00 & \multicolumn{1}{c|}{\textbf{10}} & \multicolumn{1}{c|}{0} & \multicolumn{1}{c|}{0} & \multicolumn{1}{c|}{0} & \multicolumn{1}{c|}{0} & \multicolumn{1}{c|}{0} & \multicolumn{1}{c|}{0} & \multicolumn{1}{c|}{0} & \multicolumn{1}{c|}{0} & 0 \\ \hline
Probe 2 & \multicolumn{1}{c|}{0.32} & \multicolumn{1}{c|}{\textbf{0.33}} & \multicolumn{1}{c|}{0.24} & \multicolumn{1}{c|}{0.09} & \multicolumn{1}{c|}{0.01} & \multicolumn{1}{c|}{0.02} & \multicolumn{1}{c|}{0.04} & \multicolumn{1}{c|}{0.03} & \multicolumn{1}{c|}{0.01} & 0.02 & \multicolumn{1}{c|}{0} & \multicolumn{1}{c|}{\textbf{10}} & \multicolumn{1}{c|}{0} & \multicolumn{1}{c|}{0} & \multicolumn{1}{c|}{0} & \multicolumn{1}{c|}{0} & \multicolumn{1}{c|}{0} & \multicolumn{1}{c|}{0} & \multicolumn{1}{c|}{0} & 0 \\ \hline
Probe 3 & \multicolumn{1}{c|}{0.12} & \multicolumn{1}{c|}{0.25} & \multicolumn{1}{c|}{\textbf{0.30}} & \multicolumn{1}{c|}{0.21} & \multicolumn{1}{c|}{0.06} & \multicolumn{1}{c|}{0.01} & \multicolumn{1}{c|}{0.03} & \multicolumn{1}{c|}{0.04} & \multicolumn{1}{c|}{0.02} & 0.01 & \multicolumn{1}{c|}{0} & \multicolumn{1}{c|}{0} & \multicolumn{1}{c|}{\textbf{10}} & \multicolumn{1}{c|}{0} & \multicolumn{1}{c|}{0} & \multicolumn{1}{c|}{0} & \multicolumn{1}{c|}{0} & \multicolumn{1}{c|}{0} & \multicolumn{1}{c|}{0} & 0 \\ \hline
Probe 4 & \multicolumn{1}{c|}{0.06} & \multicolumn{1}{c|}{0.13} & \multicolumn{1}{c|}{0.23} & \multicolumn{1}{c|}{\textbf{0.30}} & \multicolumn{1}{c|}{0.24} & \multicolumn{1}{c|}{0.10} & \multicolumn{1}{c|}{0.03} & \multicolumn{1}{c|}{0.01} & \multicolumn{1}{c|}{0.01} & 0.00 & \multicolumn{1}{c|}{0} & \multicolumn{1}{c|}{0} & \multicolumn{1}{c|}{0} & \multicolumn{1}{c|}{\textbf{10}} & \multicolumn{1}{c|}{0} & \multicolumn{1}{c|}{0} & \multicolumn{1}{c|}{0} & \multicolumn{1}{c|}{0} & \multicolumn{1}{c|}{0} & 0 \\ \hline
Probe 5 & \multicolumn{1}{c|}{0.01} & \multicolumn{1}{c|}{0.03} & \multicolumn{1}{c|}{0.10} & \multicolumn{1}{c|}{0.22} & \multicolumn{1}{c|}{\textbf{0.30}} & \multicolumn{1}{c|}{0.25} & \multicolumn{1}{c|}{0.12} & \multicolumn{1}{c|}{0.03} & \multicolumn{1}{c|}{0.03} & 0.04 & \multicolumn{1}{c|}{0} & \multicolumn{1}{c|}{0} & \multicolumn{1}{c|}{0} & \multicolumn{1}{c|}{0} & \multicolumn{1}{c|}{\textbf{10}} & \multicolumn{1}{c|}{0} & \multicolumn{1}{c|}{0} & \multicolumn{1}{c|}{0} & \multicolumn{1}{c|}{0} & 0 \\ \hline
Probe 6 & \multicolumn{1}{c|}{0.02} & \multicolumn{1}{c|}{0.01} & \multicolumn{1}{c|}{0.01} & \multicolumn{1}{c|}{0.10} & \multicolumn{1}{c|}{0.23} & \multicolumn{1}{c|}{\textbf{0.27}} & \multicolumn{1}{c|}{0.18} & \multicolumn{1}{c|}{0.06} & \multicolumn{1}{c|}{0.00} & 0.03 & \multicolumn{1}{c|}{0} & \multicolumn{1}{c|}{0} & \multicolumn{1}{c|}{0} & \multicolumn{1}{c|}{0} & \multicolumn{1}{c|}{0} & \multicolumn{1}{c|}{\textbf{10}} & \multicolumn{1}{c|}{0} & \multicolumn{1}{c|}{0} & \multicolumn{1}{c|}{0} & 0 \\ \hline
Probe 7 & \multicolumn{1}{c|}{0.01} & \multicolumn{1}{c|}{0.02} & \multicolumn{1}{c|}{0.03} & \multicolumn{1}{c|}{0.03} & \multicolumn{1}{c|}{0.07} & \multicolumn{1}{c|}{0.17} & \multicolumn{1}{c|}{\textbf{0.24}} & \multicolumn{1}{c|}{0.23} & \multicolumn{1}{c|}{0.12} & 0.02 & \multicolumn{1}{c|}{0} & \multicolumn{1}{c|}{0} & \multicolumn{1}{c|}{0} & \multicolumn{1}{c|}{0} & \multicolumn{1}{c|}{0} & \multicolumn{1}{c|}{0} & \multicolumn{1}{c|}{\textbf{10}} & \multicolumn{1}{c|}{0} & \multicolumn{1}{c|}{0} & 0 \\ \hline
Probe 8 & \multicolumn{1}{c|}{0.01} & \multicolumn{1}{c|}{0.02} & \multicolumn{1}{c|}{0.02} & \multicolumn{1}{c|}{0.01} & \multicolumn{1}{c|}{0.03} & \multicolumn{1}{c|}{0.09} & \multicolumn{1}{c|}{0.19} & \multicolumn{1}{c|}{\textbf{0.28}} & \multicolumn{1}{c|}{0.27} & 0.14 & \multicolumn{1}{c|}{0} & \multicolumn{1}{c|}{0} & \multicolumn{1}{c|}{0} & \multicolumn{1}{c|}{0} & \multicolumn{1}{c|}{0} & \multicolumn{1}{c|}{0} & \multicolumn{1}{c|}{0} & \multicolumn{1}{c|}{\textbf{10}} & \multicolumn{1}{c|}{0} & 0 \\ \hline
Probe 9 & \multicolumn{1}{c|}{0.01} & \multicolumn{1}{c|}{0.01} & \multicolumn{1}{c|}{0.02} & \multicolumn{1}{c|}{0.02} & \multicolumn{1}{c|}{0.01} & \multicolumn{1}{c|}{0.05} & \multicolumn{1}{c|}{0.14} & \multicolumn{1}{c|}{0.26} & \multicolumn{1}{c|}{\textbf{0.32}} & 0.30 & \multicolumn{1}{c|}{0} & \multicolumn{1}{c|}{0} & \multicolumn{1}{c|}{0} & \multicolumn{1}{c|}{0} & \multicolumn{1}{c|}{0} & \multicolumn{1}{c|}{0} & \multicolumn{1}{c|}{0} & \multicolumn{1}{c|}{0} & \multicolumn{1}{c|}{\textbf{10}} & 0 \\ \hline
Probe 10 & \multicolumn{1}{c|}{0.00} & \multicolumn{1}{c|}{0.00} & \multicolumn{1}{c|}{0.00} & \multicolumn{1}{c|}{0.00} & \multicolumn{1}{c|}{0.01} & \multicolumn{1}{c|}{0.01} & \multicolumn{1}{c|}{0.00} & \multicolumn{1}{c|}{0.06} & \multicolumn{1}{c|}{0.23} & \textbf{0.44} & \multicolumn{1}{c|}{0} & \multicolumn{1}{c|}{0} & \multicolumn{1}{c|}{0} & \multicolumn{1}{c|}{0} & \multicolumn{1}{c|}{0} & \multicolumn{1}{c|}{0} & \multicolumn{1}{c|}{0} & \multicolumn{1}{c|}{0} & \multicolumn{1}{c|}{0} & \textbf{10} \\ \hline
\end{tabular}
\label{tab5}\vspace{-0.4cm}
\end{table*}

To experimentally validate the DOA estimation performance attained, Gao \emph{et al.}~\cite{LSA_2024_Gao_Super} implemented a SIM operating at $27.5$ GHz. As illustrated in Fig.~\ref{fig_11}(a), the experimental configuration utilized a vector network analyzer (VNA) connected to horn antennas serving as target sources and a waveguide probe for signal detection. The detection region of the waveguide probe was precisely controlled via a translation stage, while a four-layer passive SIM architecture was designed and fabricated to estimate azimuth angles with $1^{\circ}$ angular resolution over the range $\left [ -5^{\circ},5^{\circ} \right ]$. Each metasurface layer consisted of a $32 \times 32$ array of modulation elements with half-wavelength spacing of $5.45$ mm. The output plane was partitioned into ten distinct detection regions, each mapped to a specific input angular range. Experimental validation using $100$ test samples yielded energy distributions and confusion matrices presented in Tab. \ref{tab5}, demonstrating that energy is concentrated at the probe corresponding to the angular interval of the incident wave. Through temporal multiplexing of different SIM configurations with varying angular resolutions and coverage ranges, this approach enables precise DOA estimation of multiple targets across an extended field of view~\cite{LSA_2024_Gao_Super}.

Additionally, SIMs are capable of functioning as electromagnetic signal processors, holding considerable promise for sensing applications with stringent latency requirements \cite{ICC_2024_An_Stacked, JSAC_2024_An_Two}. By deploying a well-trained SIM in front of a receiving array, incident electromagnetic waves can be automatically transformed into the spatial frequency domain as they propagate through the optimized SIM structure~\cite{JSAC_2024_An_Two}. Consequently, the target DOA can be directly inferred by examining the signal intensity distribution across the receiving array. To further enhance DOA estimation accuracy, An \emph{et al.}~\cite{JSAC_2024_An_Two} proposed aggregating the spatial frequency spectra acquired over multiple snapshots. As illustrated in Fig.~\ref{fig_11}(b), the transmission coefficients of the input metasurface layer were systematically reconfigured across snapshots to synthesize a series of spatial spectra with orthogonal frequency bins. Fig.~\ref{fig_10} compares the spatial frequency spectra yielded by two-layer and four-layer SIM architectures in dual-target scenarios, demonstrating that deeper SIM configurations produce a substantially more refined spatial frequency spectrum. Experimental validation further confirmed that the SIM, operating at optical computational speeds, achieves an MSE of $-40$ dB for 2D DOA estimation~\cite{JSAC_2024_An_Two}.

Moreover, the architecture of SIM can be scaled to enhance the capability in multi-level diffractive modulation of incident electromagnetic fields~\cite{WC_2024_An_Stacked}. A larger number of meta-atoms per layer translates to an expanded aperture, thereby enabling higher angular resolution in signal detection. Furthermore, increasing the network depth through additional layers allows for the estimation of a substantially higher number of target sources. Unlike conventional angle estimation techniques that impose substantial computational burdens on electronic systems, a SIM performs DOA estimation at the speed of light, making it particularly advantageous for latency-sensitive applications, such as autonomous vehicles and high-speed rail communications. Additionally, when multiple sources are present, the input electromagnetic field comprises a superposition of plane waves arriving from distinct directions. As a result, the DOA estimates are subsequently obtained by identifying the top-$K$ intensity peaks across all detection regions, where $K$ denotes the number of incident signals. Notably, this detection mechanism inherently enables the designed SIM architecture to simultaneously estimate both the angles and the number of target sources~\cite{LSA_2024_Gao_Super}.

\section{SIM for Computing}\label{sec6}
SIMs have a transformative impact on future computing architectures, opening up new avenues for integrating computing functions into the wireless transmission process. Next, a pair of representative examples are provided for demonstrating its exciting potential way beyond wireless communication and sensing applications.

\subsection{Pattern Generation}
Computational imaging is capable of overcoming the limitations of traditional optical imaging techniques~\cite{CI_2018_Mait_Advances}. By leveraging a SIM as a front-end, novel imaging capabilities can be achieved by precisely controlling its internal electromagnetic fields and pairing it with appropriate computational algorithms. Building upon this philosophy, Hassan \emph{et al.}~\cite{OJCOMS_2024_Hassan_Efficient} established an analytical framework for calculating the received power of signals after passing through a SIM. They further proposed a pair of distinct optimization approaches: a GD method for meta-atoms featuring continuous phase modulation, and a successive refinement technique for those with discrete phase configurations. Their simulation results revealed that a three-layer SIM capable of continuous phase adjustment could concentrate over $90$\% of radiated power within targeted regions. By contrast, achieving comparable power concentration with binary phase shift meta-atoms necessitated more than five metasurface layers.

To address the substantial challenge of high maintenance costs associated with conventional image storage, Fan \emph{et al.}~\cite{NC_2024_Fan_Holographic} introduced the concept of metasurface-based disks (meta-disks) that significantly extend the capacity limits of holographic storage by exploiting uncorrelated structural twist angles. By utilizing these independent spatial DoFs, this approach enables multi-dimensional structural multiplexing, exponentially increasing the data storage density compared to traditional single-layer holographic media without requiring complex optical setups. Utilizing advanced and cost-effective 3D printing technology combined with Pancharatnam-Berry metasurfaces \cite{NC_2024_Fan_Holographic}, the authors fabricated a twin-layer SIM architecture with an interlayer spacing of $1,000$ $\mu$m, where each layer comprises circular disks containing $256$ meta-atoms in radius. This additive manufacturing approach drastically minimizes fabrication expenses, offering a highly accessible and economically viable hardware solution for mass production. While a slight energy consumption is inherently required for the mechanical rotation of these metasurfaces, the fundamental performance versus hardware trade-offs of this system warrant further clarification in future design optimizations.

\begin{figure}[!t]
\centering
\includegraphics[width=8.5cm]{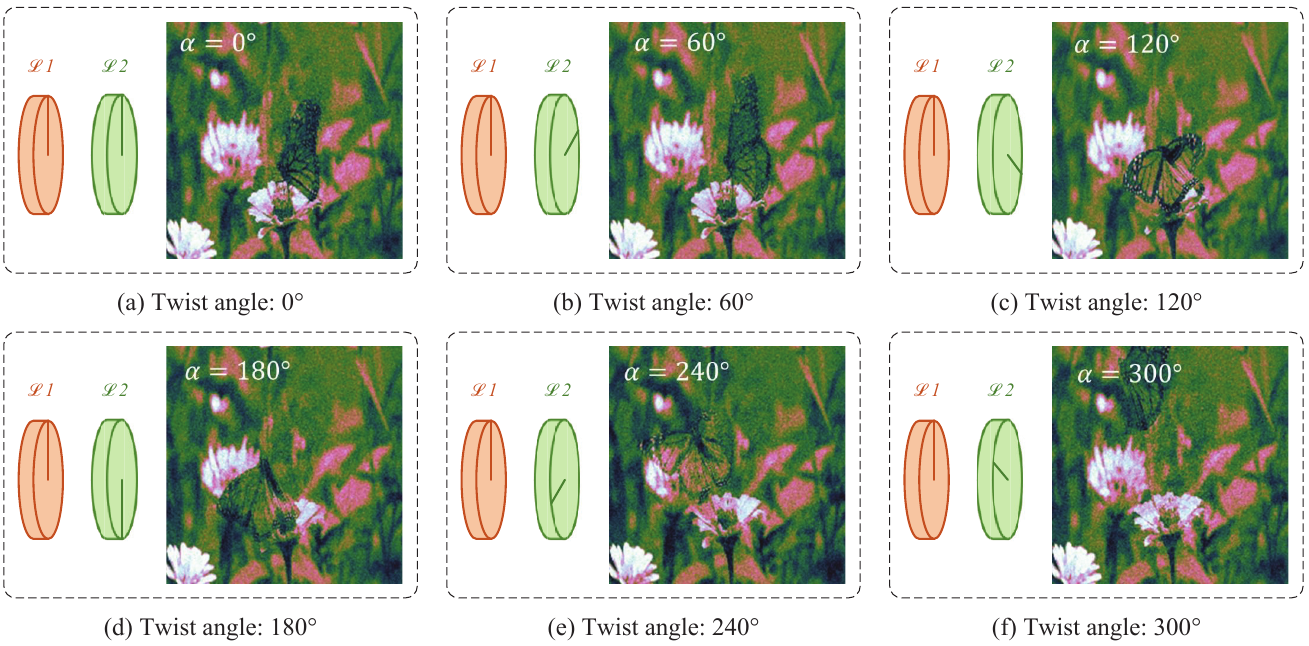}
\caption{Holograms generated by a two-layer SIM at six different twist angles.}
\label{fig_12}\vspace{-0.4cm}
\end{figure}

To demonstrate the SIM's potential for holographic display applications, the temporal information from an original video was encoded into different twist angles between the two meta-disk layers. As illustrated in Fig.~\ref{fig_12}, high-quality video sequences with a resolution of $200$ $\times$ $200$ pixels were successfully reconstructed at the designated output plane by sequentially adjusting the metasurface twist angle at intervals of $60^{\circ}$. Nonetheless, the fundamental storage capacity limit is governed by an inherent trade-off between the metasurface's physical size and the angular resolution of the structural multiplexing. Given a full $360^{\circ}$ rotational range and quantizing a video sequence into multiple discrete angular sectors, each step interval corresponds to a single frame. As the number of frames increases, the angular separation decreases, which inherently heightens the structural correlation between adjacent states and exacerbates image crosstalk during reconstruction. By scaling up the aperture of the metasurface, the correlation between closely spaced angular states can be further reduced, thereby suppressing crosstalk and significantly improving the fidelity of the reconstructed images.

This advanced pattern generation technology holds significant promise for next-generation multiple access using holograms and may potentially unlock new possibilities for generative AI applications. Moreover, this method can be utilized to enable semantic information representation, particularly for intricate scene information such as geometric configurations and spatial arrangements, which are difficult to describe accurately using conventional techniques. By using a SIM to directly generate the desired shape and edge at the receiving array, the amount of data traffic could be tremendously reduced~\cite{WCL_2025_Huang_Stacked}.

\subsection{Logical Operation}
In addition, SIMs offer promising capabilities for implementing logical operations in the spatial domain, which holds substantial potential for future bit-level electromagnetic information processing. A recent experimental demonstration~\cite{LSA_2020_Qian_Performing} showcased a twin-layer microwave-frequency SIM that successfully realized the three fundamental binary logical operations: AND, OR, and NOT. The architecture employs an input layer comprising a patterned mask with multiple switchable regions, each capable of toggling between high and low transmittance states, where the high transmittance state designates regions activated for logical computation. At the output plane, two distinct regions encode the binary logic states `1' and `0', respectively. Through training the phase modulation of each hidden layer via error backpropagation~\cite{LSA_2020_Qian_Performing}, the SIM achieves precise electromagnetic wave routing to the appropriate output region. Building upon this foundation, Ding \emph{et al.}~\cite{AM_2024_Ding_Metasurface} introduced a novel architecture for optical quantum-domain logical operations leveraging spatial and polarization multiplexing within a SIM framework. In their approach, the pair of quantum states, $\left| 0 \right>$ and $\left| 1 \right>$, are encoded using orthogonal linear polarization states of classical electromagnetic waves. By superimposing these basis states, the optimized SIM successfully executes four fundamental quantum-domain logical operations: the three Pauli gates and the Hadamard gate. Remarkably, the optical quantum operator implemented exhibited exceptional fidelity across all four gates, achieving $99.96\%$ in numerical simulations and $99.88\%$ in experimental validation.

\begin{figure}[!t]
\centering
\includegraphics[width=8.5cm]{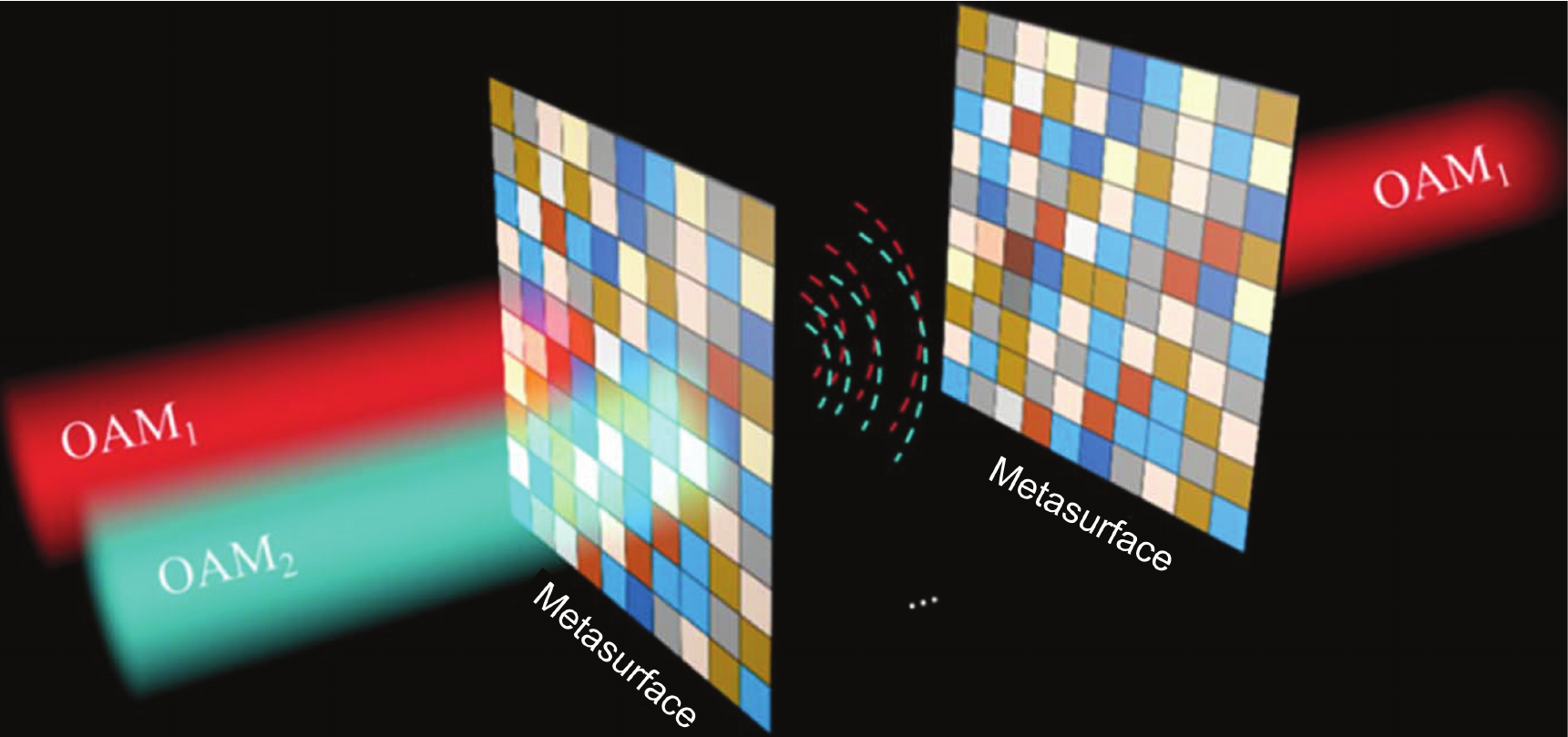}
\caption{Schematic of OAM mode logical operation based on SIM \cite{PR_2021_Wang_Orbital}.}
\label{fig_14}
\end{figure}

\begin{figure}[!t]
\centering
\includegraphics[width=8.5cm]{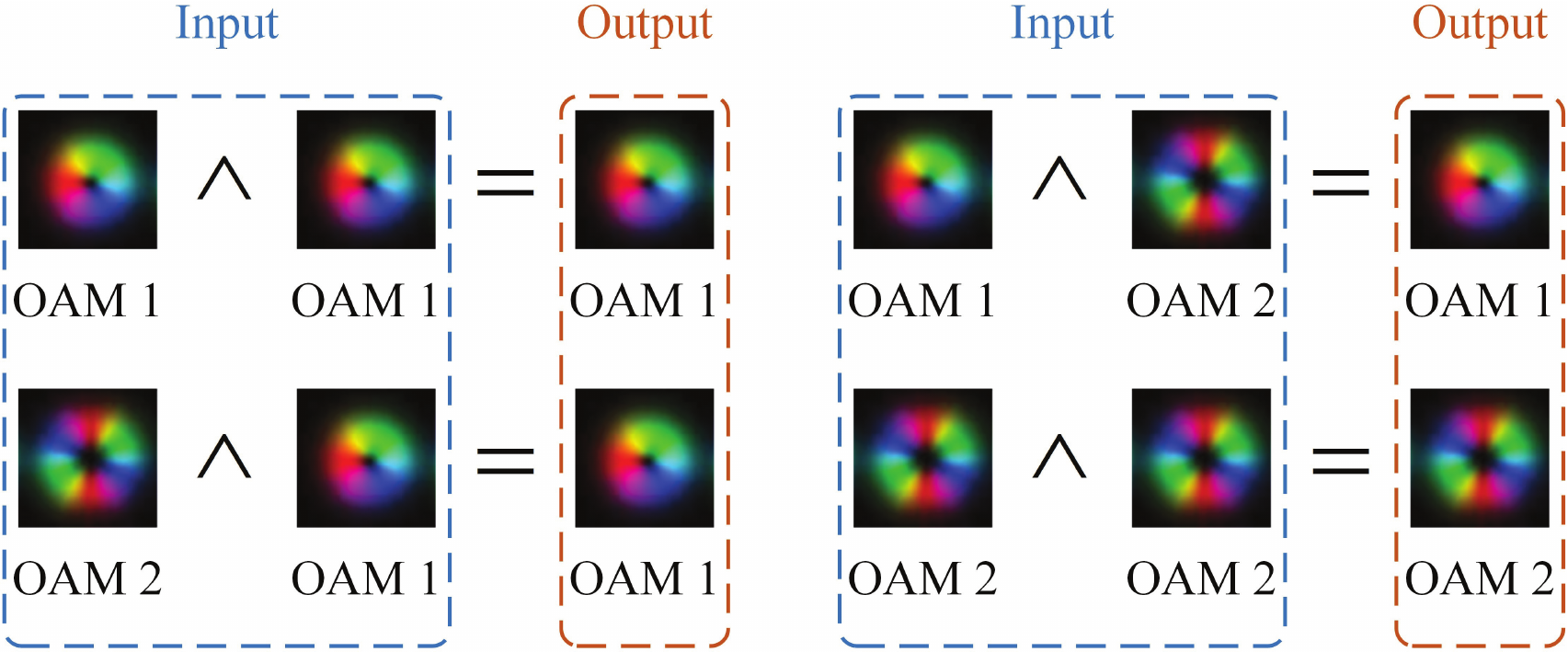}
\caption{Illustration of the logic AND gate.}
\label{fig_15}\vspace{-0.4cm}
\end{figure}

Instead of employing linear momentum states for logic representation that may suffer from ambiguous discrimination boundaries, the authors of~\cite{PR_2021_Wang_Orbital} introduced an innovative technique of leveraging OAM modes as logical states. This methodology not only augments the parallel processing capacity attained but also significantly enhances the distinguishability and robustness of logical states, capitalizing on the infinite dimensionality and orthogonality of OAM modes. Through precise manipulation of the phase and amplitude distributions across multiple diffractive layers, the SIM enables independent control of both the modal content and spatial positioning of multiple OAM beams. As illustrated in Fig.~\ref{fig_14}, two OAM modes incident from distinct spatial locations serve as input data, and the primary objective of the SIM is to perform independent modulation of these input fields, thus ensuring that the transformations conform to the computational requirements of logical gates. Following wavefront modulation through multiple diffraction screens and subsequent propagation, a vortex beam carrying a single OAM mode emerges at the center of the output plane. Fig.~\ref{fig_15} illustrates the OAM-based logic AND gate, where the first and second columns display the intensity distributions of the input beams, with the color gradient from blue to red indicating the OAM mode order. The corresponding outputs are shown in the third column of Fig.~\ref{fig_15}, which conform to that of the logic AND gate.

\begin{figure}[!t]
\centering
\includegraphics[width=8cm]{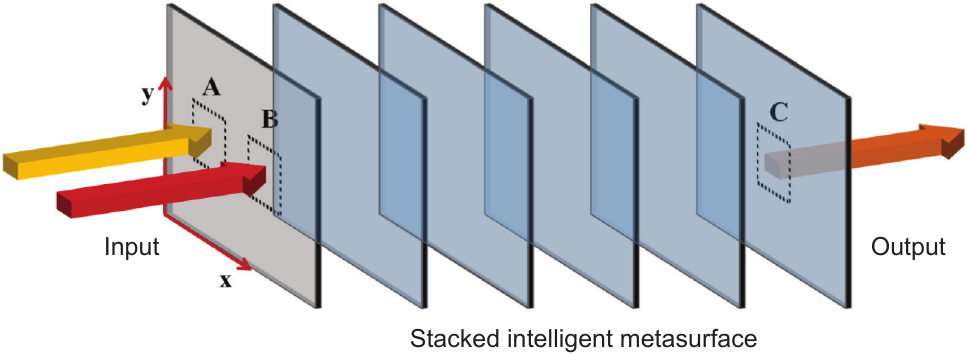}
\caption{A schematic of logic gates using polarization DoF based on SIM. Ports A and B on the input layer represent the two inputs, while C represents the output port located at the center of the output layer \cite{JO_2024_Lin_Polarization}.}
\label{fig_13}\vspace{-0.4cm}
\end{figure}

In~\cite{NP_2024_Liu_All}, Liu \emph{et al.} developed a SIM architecture that achieves remarkable neuronal density for optical computing, integrating $40,000$ neurons within a compact $2\ \text{cm} \times 2\ \text{cm}$ footprint. To validate this platform, they numerically simulated an optical half-adder employing two metasurface layers and subsequently verified it experimentally in the Terahertz regime. Similarly, the authors of~\cite{JO_2024_Lin_Polarization} utilized SIM to realize a polarization-encoded logical operation exhibiting high extinction ratios. Their approach harnesses free-space electromagnetic wave propagation and sophisticated engineered interactions with cascaded passive diffractive layers to enable efficient optical computation. As depicted in Fig.~\ref{fig_13}, they implemented a five-layer architecture capable of executing various logic operations, with constituent pixels featuring dimensions on the order of several hundred nanometers. The binary logical states `0' and `1' are encoded using linearly polarized light beams with electric field orientations along the $X$ and $Y$ directions, respectively, which are injected into the input ports A and B. Upon propagation through the appropriately configured SIM, a linearly polarized beam emerges at output port C, corresponding to the logical states `0' or `1', respectively. By enriching the input encoding patterns and adopting more sophisticated network architectures, SIMs can theoretically realize arbitrary logical functions.

Despite its immense potential, the end-to-end implementation of secure, hardware-level encryption pipelines via SIM architectures remains hindered by several formidable challenges. Chief among these is the architectural complexity introduced by intricate cryptographic protocols, which necessitate deeply cascaded SIM topologies. Such multi-layered configurations exponentially complicate both upper-layer network protocol design and spatial wave-routing control. From a physical layer perspective, a major bottleneck arises from intense electromagnetic mutual coupling and inter-SIM interference within tightly spaced cascades, which can significantly compromise signal fidelity and mapping accuracy. Concurrently resolving these hardware and architectural trade-offs demands holistic future investigations into joint layer-protocol co-design and robust mutual coupling mitigation frameworks.
\section{Fusion of Communication, Sensing, and Computing}
Emerging application scenarios increasingly necessitate the convergence of communication, sensing, and computing functionalities within a unified infrastructure. The integration of radar sensing and edge computing capabilities into WiFi and cellular networks, for example, demonstrates considerable potential for efficient radio resource orchestration~\cite{Arxiv_2024_Wang_Multi, WC_2024_An_Stacked, OJCOMS_2025_Magbool_A}. This convergence paradigm suggests that a single multifunctional SIM device could simultaneously fulfill these diverse objectives~\cite{arXiv_2025_Fadakar_Stacked}. In this section, we examine existing advances in multiplexing SIM technology to accomplish multiple concurrent tasks.

\subsection{Algorithmic Frameworks}
In~\cite{WCL_2024_Niu_Stacked}, Niu \emph{et al.} explored a SIM-enhanced ISAC framework, wherein the SIM was configured to produce a tailored beam pattern enabling simultaneous multi-user downlink communication and radar target detection. To achieve this end, the authors formulated an optimization problem aimed at maximizing spectral efficiency subject to directional power constraints, necessitating the joint optimization of the SIM phase shifts and the BS's power allocation. By treating the sensing power constraint as a penalty term within the objective function, they developed and solved the resultant optimization problem using a customized GA algorithm. Following this, Li \emph{et al.}~\cite{arXiv_2024_Shunyu_Transmit} conceived an advanced transmit beamforming technique for SIM-enabled ISAC systems, where they employed a multi-layer beamformer design that maximizes user sum rate, while optimally sculpting the normalized beam pattern for sensing purposes. The resultant non-convex multi-objective optimization problem was addressed through a dual-normalized differential GD algorithm, which leverages gradient differences and dual normalization to strike a flexible trade-off between communication and sensing performance objectives. Furthermore, Jiang \emph{et al.}~\cite{WCL_2025_Jiang_Stacked} designed a satellite equipped with SIM to perform dual-function precoding in the wave domain — eliminating the need for conventional digital beamformers — for simultaneously detecting multiple airborne targets, while serving multiple ground-based communication users. A penalty-based GA algorithm was developed for maximizing the communication throughput attained, while maintaining sensing performance.

\begin{figure}[!t]
\centering
\includegraphics[width=8cm]{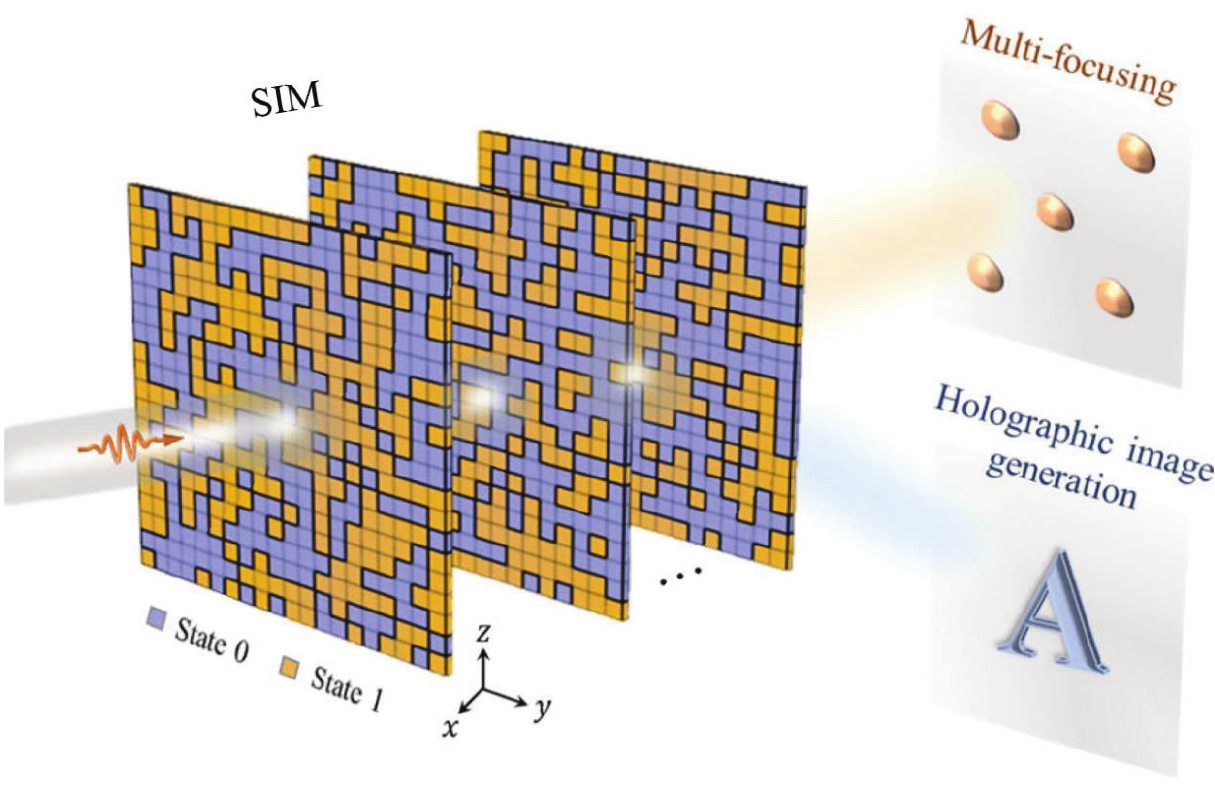}
\caption{A schematic showing the dual-functional capability of SIM for achieving multifocusing and holographic imaging \cite{TAP_2024_Jia_High}.}
\label{fig_16}\vspace{-0.4cm}
\end{figure}

Recently, Gao \emph{et al.}~\cite{LSA_2024_Gao_Super} have demonstrated the significant potential of SIM in RIS-aided mmWave ISAC systems. In contrast to conventional RIS-based communication systems, where DOA estimation necessitates for the BS to execute a series of computationally intensive operations, including down-conversion, sampling, and digital signal processing, a SIM directly infers angular information from the incident signals and facilitates edge computing. This capability substantially reduces both the power consumption and latency in beam-tracking applications. Specifically, they developed a reflective liquid crystal (LC)-based RIS architecture comprising $20 \times 20$ programmable meta-atoms, where each meta-atom achieves $5$-bit phase modulation of the incident electromagnetic field through FPGA-controlled supply voltages. By employing a passive SIM to estimate the DOAs of signals originating from both the BS and user, the RIS dynamically optimizes its beamforming configuration to establish a virtual communication link with high-velocity mobile users. Experimental results reveal that the system achieves an average amplitude gain of $17.9$ dB at the user's antenna compared to configurations operating without SIM~\cite{LSA_2024_Gao_Super}.

Moreover, Jia \emph{et al.}~\cite{TAP_2024_Jia_High} explored the potential of using a SIM to support the dual functions of imaging and beamforming. Specifically, they developed an advanced architecture employing two layers of tunable binary transmissive metasurfaces that are capable of generating complex holographic patterns on a 2D plane. By optimizing the binary phase of each meta-atom to minimize discrepancies between the actual and desired radiation patterns, as illustrated in Fig.~\ref{fig_16}, this SIM configuration can simultaneously achieve multi-focusing and holographic imaging capabilities, while maintaining a transmission efficiency exceeding $95$\%. Beyond the aforementioned applications, SIMs also hold considerable potential for enabling covert communications. By steering a deep null toward a designated eavesdropping target while simultaneously suppressing multi-user interference, the probability of communication interception can be substantially reduced. Nevertheless, this application domain remains in its nascent stages, and several open challenges — encompassing physical layer protocol design, beamforming algorithm development, and covertness performance analysis — warrant thorough investigation in future research.

\begin{figure}[!t]
\centering
\includegraphics[width=8.5cm]{ISAC_experiment.jpg}
\caption{The experiment on the application of SIM in ISAC systems \cite{Arxiv_2024_Wang_Multi}.}
\label{fig:ISAC}\vspace{-0.4cm}
\end{figure}
\subsection{Experimental Validation}
To empirically validate the viability of wave-domain signal processing for dual-functional objectives, the authors of \cite{Arxiv_2024_Wang_Multi} developed a multi-user ISAC system driven by an X-band transmissive SIM hardware prototype. As illustrated in Fig. \ref{fig:ISAC}(a), the physical architecture employs a cascaded configuration of two transmissive metasurface layers, each populated by a $16 \times 16$ meta-atom array. Each constituent meta-atom is engineered using a multi-layer microstrip substrate integrated with a commercial surface-mount PIN diode, which provides 1-bit digital phase quantization while maintaining a stable, low insertion loss profile. Over-the-air validation was carried out inside a fully shielded microwave anechoic chamber. In this experimental setup, a conventional horn antenna serves as the BS feeder array to illuminate the SIM's input layer, several forward-facing horn antennas act as single-antenna communication users to assess downlink signal-to-interference-plus-noise ratio (SINR) enhancements, and a discrete scattering object is positioned within the target sector to evaluate the sensing Cramér-Rao Bound (CRB) via an auxiliary receiver array. Ultimately, the empirical findings demonstrate that the semidefinite relaxation (SDR)-based AO frameworks successfully orchestrate real-time, wave-domain analog beamfocusing. This prototype conclusively proves that SIMs can isolate multi-user communication streams while simultaneously synthesizing directed sensing beams entirely within the electromagnetic propagation domain, thereby bypassing the need for computationally heavy baseband digital precoding networks.

\begin{table*}[!t]
\centering
\renewcommand\arraystretch{1.5}
\caption{A survey of existing channel estimation schemes for SIM-based communication systems.}
\begin{tabular}{l|c|c|c|c|l}
\hline
Reference & Year & Link & Scenario & Estimator & Contributions \\ \hline
Nadeem \emph{et al.}~\cite{WCNC_2024_Nadeem_Hybrid} & 2024 & UL & MU & MMSE estimator & Leverage GD algorithm to configure the SIM for minimizing MSE \\ \hline
Yao \emph{et al.}~\cite{WCL_2024_Yao_Channel} & 2024 & UL & MU & Subspace-based estimators & \begin{tabular}[c]{@{}l@{}}Utilize the low-rank characteristics of the channel subspace to improve\\the estimation performance\end{tabular} \\ \hline
Yao \emph{et al.}~\cite{APWCS_2024_Yao_Sparse} & 2024 & UL & MU & OMP & Utilize the decision mechanism to adaptively determine the sparsity \\ \hline
Ginige \emph{et al.}~\cite{CL_2025_Ginige_Nested} & 2025 & UL & MU & Alternative LS & \begin{tabular}[c]{@{}l@{}}Use PARAFAC and Tucker methods to represent multidimensional\\algebraic structures\end{tabular} \\ \hline
Lawal \emph{et al.}~\cite{CL_2025_Lawal_Channel} & 2025 & UL & MU & CNN & Learn the internal propagation behavior without explicit physical modeling \\ \hline
\end{tabular}\label{tab6}\vspace{-0.4cm}
\end{table*}

Furthermore, the empirical evaluations in \cite{Arxiv_2024_Wang_Multi} have clarified a pivotal design principle regarding the interplay between the received power gain, the inter-layer spacing, and the overall layer count, thereby resolving the misconception that scaling the number of layers inherently compromises system gain. As illustrated in Fig. 7(b), when the inter-layer spacing is constrained to half a wavelength ($\le \lambda/2$) or less, the signal successfully propagates through successive layers; despite the accumulation of penetration losses, the substantial aperture area of the SIM yields a power gain that effectively compensates for these impairments, causing the received power to scale monotonically with the number of layers. Conversely, this trend reverses once the inter-layer spacing exceeds a certain value. In this regime, the expanded propagation distance significantly exacerbates inter-layer attenuation, which ultimately eclipses the aperture gain of the metasurface. Consequently, these findings provide a vital deployment guideline, indicating that multi-layer SIM architectures need to maintain a highly compact physical topology to minimize internal electromagnetic attenuation.
\section{Key Technical Issues Related to SIMs}
In this section, we explore the practical challenges of implementing SIMs. Specifically, we cover several key aspects, including channel estimation, antenna selection and user association, inter-layer propagation coefficient calibration, propagation modeling, and EE analysis.

\subsection{Channel Estimation}
Accurate channel state information (CSI) is vital for carrying out the desired signal processing function in the wave domain, particularly in dynamic wireless communication systems~\cite{WC_2024_An_Stacked, TVT_2026_Dong_Deep}. Typically, each metasurface layer consists of a large number of meta-atoms, which leads to a situation where the number of RF chains at the BS is much lower than the dimension of channels between the SIM and users. This disparity renders CSI acquisition in SIM-assisted communication systems an under-determined problem. To tackle this issue, Nadeem \emph{et al.}~\cite{WCNC_2024_Nadeem_Hybrid} developed a channel estimation technique that collects multiple copies of uplink pilot signals that propagate through the SIM. Orthogonal pilot signals were employed to prevent interference between users. Moreover, they theoretically analyzed the MSE of the channel estimates and applied a GD algorithm to optimize the phase shifts of meta-atoms, aiming for minimizing the MSE. Notably, they also demonstrated that by leveraging the low-rank property of channel correlation matrices~\cite{CL_2023_An_A}, the pilot overhead can be significantly reduced.

Furthermore, Yao \emph{et al.}~\cite{WCL_2024_Yao_Channel} identified a subspace that spans any spatial correlation matrix when ultra-dense elements are arranged on a metasurface. By relying on realistic partial CSI of the channel statistics, two subspace-based channel estimators were developed for enhancing the estimation accuracy by projecting both conventional least squares (LS) and MMSE estimates into the lower-dimensional subspace they identified. In~\cite{APWCS_2024_Yao_Sparse}, the same researchers exploited the sparse nature of mmWave channels in the beam domain. They framed the channel estimation problem of SIM-aided mmWave communication systems as a sparse recovery task and applied the popular orthogonal matching pursuit (OMP) algorithm to determine the channel parameters. Additionally, they introduced a soft decision threshold for terminating the OMP iterations when the sparsity is unknown. Recent studies~\cite{CL_2025_Ginige_Nested} and~\cite{CL_2025_Lawal_Channel} have also applied alternative LS and convolutional neural networks (CNNs) to address channel estimation challenges in SIM-assisted communication systems. To illustrate, the existing channel estimation techniques conceived for SIM-assisted communication systems are summarized in Table \ref{tab6}.

\subsection{Antenna Selection and User Association}
\begin{figure}[!t]
\centering
\includegraphics[width=8.5cm]{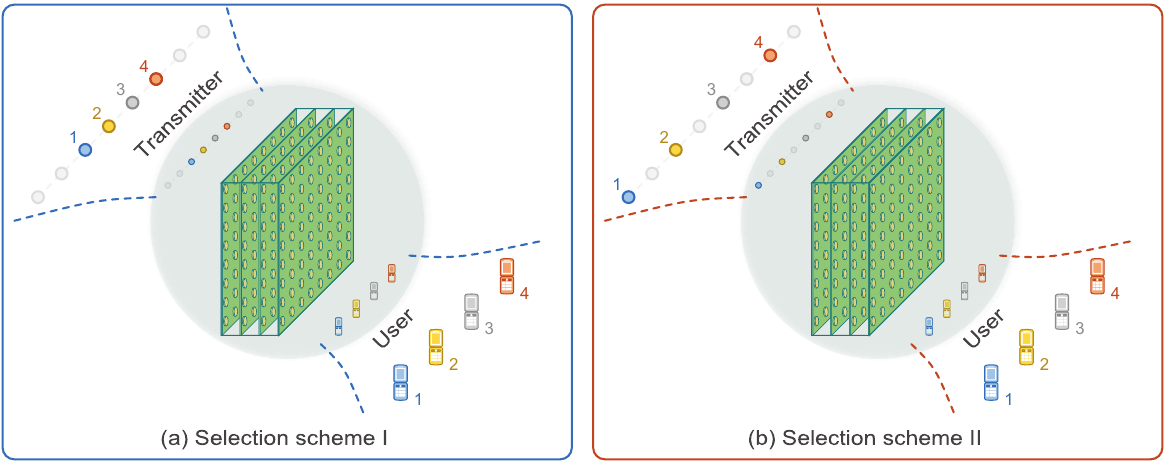}
\caption{Illustration of two different antenna selection schemes.}
\label{fig_17}
\end{figure}

\begin{figure}[!t]
\centering
\includegraphics[width=8.5cm]{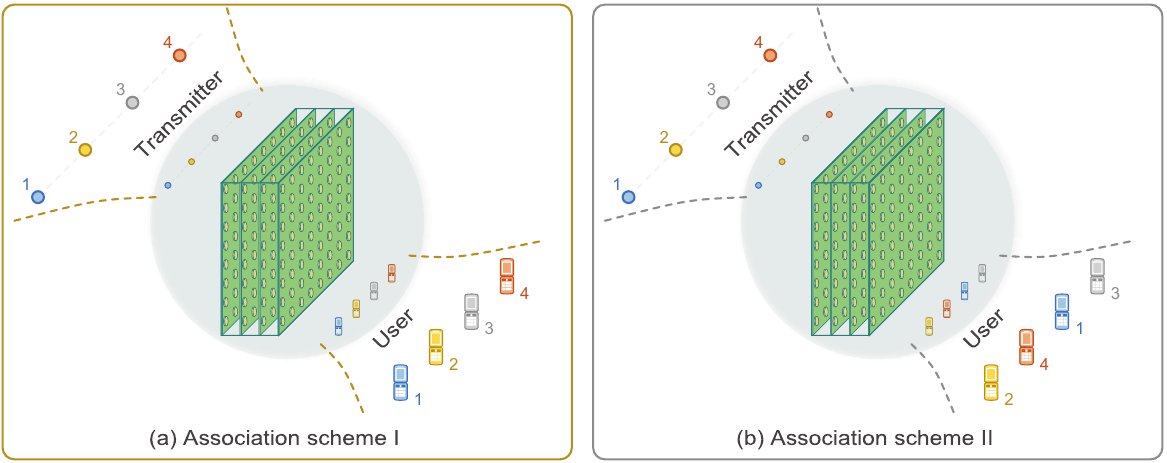}
\caption{Illustration of two different user association schemes.}
\label{fig_18}\vspace{-0.4cm}
\end{figure}

SIM essentially performs specific transformations, as waves propagate through it. For SIM to work effectively, it is crucial to establish bespoke protocols that define the role of each port. The assignment and definition of these ports significantly impact the overall performance. In wireless communication systems, this involves selecting the appropriate antennas and matching them to the right users. To illustrate this concept, let's consider a multiuser transmission scenario, where a SIM is utilized to mitigate interference among multiple users in the wave domain.
\begin{itemize}
\item[$\diamond$] \textsl{\textbf{Antenna Selection:}} As shown in Fig.~\ref{fig_17}, different antenna selection schemes can influence the capability of SIMs for mitigating inter-user interference. When the antennas are too close to each other, it becomes harder for users to differentiate their individual signals from the superimposed waveforms. Hence, selecting antennas that are spaced far apart enhances the ability to achieve interference-free transmission using the wave-domain beamforming technique.
\item[$\diamond$] \textsl{\textbf{User Association:}} As illustrated in Fig.~\ref{fig_18}, the choice of the specific user association strategy also affects the performance of wave-domain beamforming. Following the same principle as in classic antenna selection, it is suggested that neighboring antennas at the BS should transmit signals to two users who are farther apart. This setup allows for achieving effective interference mitigation.
\end{itemize}

Specifically, when considering a scenario, where a BS with $M$ antennas serves $K$ users, the total number of possible combinations for implementing antenna selection and user association can be expressed as
\begin{align}
 P\left ( M,K \right )=\frac{M!}{\left ( M-K \right )!}.
\end{align}
Therefore, determining the optimal pairing scheme of antennas to users is an NP-hard problem, necessitating the development of efficient heuristic methods. A similar challenge exists in sensing and computing applications. For instance, in object recognition, the assignment of probes to various image categories can have a significant impact on the classification performance. As such, these assignments must be carefully planned before the practical implementation of SIM.

At the time of writing, research on antenna selection and user association in SIM-aided systems is still in its early stages. In~\cite{WCL_2024_Lin_Stacked}, Lin \emph{et al.} investigated the impact of antenna selection on SIM performance in LEO satellite communications. They developed a user grouping method to manage large numbers of users by dividing them into multiple groups based on channel correlation. Within each group, users are served using spatial division multiple access (SDMA), while different groups are served in separate time slots. To further enhance system performance, the Hungarian algorithm was used at the satellite to optimize antenna selection~\cite{WCL_2024_Lin_Stacked}. In~\cite{arXiv_2024_Enyu_Joint}, Shi \emph{et al.} examined the AP-UE association in cell-free systems. They employed a greedy algorithm to pair UEs with their nearest APs and then jointly optimized power allocation and wave-based precoding to maximize data rates.

\subsection{Propagation Coefficient Calibration}
The deformation of the mechanical support structure during the assembly of the SIM, as well as the bending of the metasurfaces due to their weight, can cause the propagation coefficients $w_{n,\tilde{n}}^{l}$ between adjacent metasurface layers to deviate from their initial values modelled in \eqref{eq_1}. This deviation presents challenges during the configuration of the SIM in the training process and may negatively impact its inference performance. Therefore, before implementing the SIM, it is crucial to appropriately calibrate the wave propagation coefficients between adjacent metasurface layers to ensure more reliable training results~\cite{arXiv_2024_Jiancheng_Emerging}. In~\cite{NE_2022_Liu_A}, Liu \emph{et al.} utilized $5,000$ experimental samples to fine-tune the propagation coefficients by employing the gradient descent method. For each sample, they applied random bias voltages to all the meta-atoms in the SIM and measured the resultant output field patterns using a custom array of $8 \times 8$ receiving antennas. After calibrating the propagation coefficients, they found that the measured energy distributions closely matched the simulated ones, demonstrating excellent agreement.

\subsection{Propagation Modeling}
Accurate propagation modeling is essential for producing the desired inference results~\cite{TWC_2025_Niu_Introducing}. Given that SIMs rely on the cascaded response of multiple metasurface layers, several recent contributions have been made to characterize the near-field wave propagation within SIMs. In~\cite{arXiv_2024_Maryam_Uplink}, Maryam \emph{et al.} investigated the achievable sum-rate of SIMs in uplink scenarios while accounting for hardware imperfections. The authors formulated a non-convex sum-rate optimization problem and employed genetic algorithms alongside interior point methods to obtain the solution. Evaluating the performance under both Rayleigh fading and 3GPP channel models under constraints on either equal numbers of RF chains or equal physical aperture sizes, they demonstrated that SIMs outperform conventional digital phased arrays, when the number of RF chains is held constant. Building upon this, Nerini \emph{et al.}~\cite{CL_2024_Nerini_Physically} developed a physically consistent channel model for SIM-aided systems that explicitly incorporates realistic element coupling effects and accounts for metasurfaces having non-diagonal phase shift matrices.

More recently, Abrardo \emph{et al.}~\cite{arXiv_2025_Andrea_A} proposed a comprehensive optimization framework for heterogeneous SIM architectures that circumvents the restrictive assumptions of prior approaches. The model they presented is grounded in multi-port network theory for characterizing general electromagnetic collaborative objects (ECOs) and establishes a unified optimization framework. Subsequently, they view SIMs as a specific ECO architecture, providing valuable insights into optimization strategies across various SIM configurations and analyzing the associated computational complexity. The authors further examine the implications of commonly adopted modeling assumptions and introduce a backpropagation-based algorithm for implementing GD optimization in simplified SIM configurations.

\subsection{Energy Efficiency Analysis}
Energy efficiency analysis is paramount for validating the low-power efficacy of SIM architectures in wave-domain signal processing applications. In \cite{arXiv_2024_Stefan_Energy}, Stefan \emph{et al.} investigated the EE performance of SIM-assisted MIMO broadcast networks. To rigorously benchmark the capabilities of the SIM, the authors evaluated both dirty paper coding (DPC) and linear precoding schemes. They formulated corresponding non-convex EE maximization frameworks and exploited broadcast channel (BC)-multiple-access channel (MAC) duality to derive tractable equivalent formulations. To solve these, the users' covariance matrices were optimized via successive convex approximation (SCA) utilizing a tight lower bound on the achievable sum-rate, coupled with Dinkelbach’s algorithm, while the high-dimensional phase-shift configuration was resolved using a projected gradient method. Subsequently, Shi \emph{et al.} \cite{TWC_2025_Shi_Energy} integrated a SIM into multi-antenna BSs for downlink multi-user communications, introducing a realistic hardware power consumption model. They employed a quadratic transform to reformulate the fractional EE objective and proposed an AO-based joint precoding framework. Specifically, an SCA algorithm was adopted for the BS digital precoder design, whereas the wave-domain analog beamforming was tackled via two distinct strategies: a high-performance semidefinite programming (SDP) approach and a computationally efficient projected gradient ascent (PGA) algorithm. Notably, their findings revealed that while the optimal layer count diverges for spectral efficiency and EE maximization objectives, a highly compact configuration of $2 \sim 5$ layers simultaneously satisfies both metrics \cite{TWC_2025_Shi_Energy}.

Despite these foundational efforts, a rigorous, universally applicable EE analysis for SIM-aided systems remains exceptionally challenging. This difficulty stems from the multi-physical complexity of modeling internal energy losses, which demands the concurrent characterization of wave-domain inter-layer propagation dynamics, the direct power dissipation of active metasurface components, and the corresponding power reduction achieved by offloading digital baseband functions. Furthermore, the performance trade-offs and cost structures associated with these design variables diverge drastically between idealized academic simulations and industrialized hardware deployments. Consequently, bridging these gaps necessitates extensive, synergistic investigations across both hardware fabrication and electromagnetic modeling paradigms.

\section{Future Research Directions}
In this section, we identify research opportunities and challenges ahead for further enhancing the inference capabilities and computational power of SIM.

\subsection{Channel Multiplexing}
\begin{figure*}[!t]
\centering
\includegraphics[width=13cm]{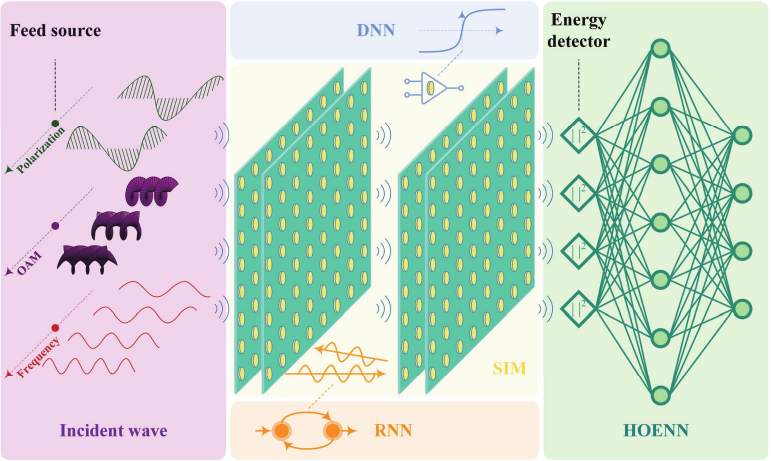}
\caption{Schematic of multifunctional SIM, where the number of available channels can be expanded by multiplexing polarization, OAM, and frequencies. In addition, by integrating nonlinear modules, leveraging inter-layer signal reflections, or cascading an electronic network, DNN, RNN, or HOENN can be created to further improve the computing power of SIM.}
\label{fig_19}\vspace{-0.4cm}
\end{figure*}

In addition to directly processing the phase and amplitude of input waves, SIMs can tap into all the DoFs exhibited by electromagnetic waves. Various multiplexing schemes, such as polarization, OAM, and frequency, can be leveraged to expand the number of available channels, as illustrated in Fig.~\ref{fig_19}.
\begin{itemize} 
 \item[$\diamond$] \textsl{\textbf{Polarization:}} The subwavelength structures of anisotropic meta-atoms that exhibit polarization-selective responses allow for independent, pixel-level control over polarization states~\cite{LSA_2022_Luo_Metasurface}.
 \item[$\diamond$] \textsl{\textbf{OAM:}} Moreover, metasurfaces bestow the unprecedented ability to manipulate structured electromagnetic waves to multiplex information in different OAM channels~\cite{Sci_2018_Lin_All}.
 \item[$\diamond$] \textsl{\textbf{Frequency:}} Sophisticated meta-atoms also provide unique opportunities for operating across multiple wavelength channels by simultaneously tuning the modulation response and by carefully engineering the native dispersion characteristics of individual meta-atoms on demand~\cite{SA_2021_Li_Spectrally}.
\end{itemize}

While these parallelization strategies offer clear advantages in scaling computational capacity and maximizing the degree of parallelism, they inherently require independent channel processing to execute concurrent multi-tasking. However, accurately modeling the unique characteristics of these distinct channels while mitigating potential inter-channel interference remains a critical open research direction. As the channel density scales within a single SIM device, multiple channels may be forced to share a highly constrained set of metasurface parameters to meet their respective signal processing requirements. This optimization strain can introduce significant performance degradation and accuracy loss. Consequently, it is essential to rigorously characterize the fundamental capacity upper bounds for SIM channel multiplexing. Furthermore, these theoretical multiplexing frameworks must be validated through more extensive experimental verification for pratical deployements.

\subsection{Structural Parameter Analysis}
Similar to an ANN, the representation capability of a SIM is influenced by the number of metasurface layers and the number of meta-atoms per layer. Additionally, the per-layer topology of meta-atoms and inter-layer spacing have an impact on the inference ability of a SIM. Understanding the influence of these hardware parameters is essential for determining the most effective configuration for a practical SIM device.
\begin{itemize}
\item[$\diamond$] \textsl{\textbf{Number of Layers:}} It has been shown that a SIM can approximate any universal linear transformations by performing cascaded complex-valued matrix operations with arbitrarily small error~\cite{Sci_2018_Lin_All}. The dimensionality of the space for linear transformations that can be processed by a SIM is linearly proportional to the number of metasurface layers. In general, deep architectures comprising a larger number of metasurface layers are capable of extracting better features compared to shallow ones. However, they also suffer from increased material absorption and surface back reflections, which drastically degrade the power efficiency. Therefore, the fundamental tradeoff between inference accuracy and power efficiency makes the use of a massive number of cascaded metasurface layers challenging.
\item[$\diamond$] \textsl{\textbf{Number of Meta-atoms:}} Employing more meta-atoms within each layer can improve the performance of a SIM. However, the law of diminishing returns holds, as the meta-atoms located far from the paraxial region have a negligible impact on the desired input-output transformations.
\item[$\diamond$] \textsl{\textbf{Per-layer Topology:}} Periodic metasurface designs may result in redundant propagation paths within a SIM. Hence, altering the topology of subwavelength meta-atoms can introduce additional flexibility to tune the transfer function of a SIM. Designing sophisticated meta-atom patterns and sparse interconnections could enhance the performance for a given number of meta-atoms or achieve desired functionalities with a more efficient SIM architecture~\cite{NE_2023_Gao_Programmable}.
\item[$\diamond$] \textsl{\textbf{Inter-layer Spacing:}} The connectivity within a SIM can also be controlled by adjusting the distance between metasurface layers using micro-electromechanical systems. Alternatively, the effective path length between metasurface layers can be tailored by tuning the refractive index of phase-changing materials without changing the physical distances.
\end{itemize}
In a nutshell, designing an effective SIM requires carefully balancing these fundamental tradeoffs, which remains an open challenge.

\subsection{Neural Network Architecture}
The concept of neural networks has been utilized for explaining the operating principles of SIMs. However, a SIM does not inherently include any non-linearity, except for the receiving detector. This limitation may degrade the ability of SIM to extract complex features and handle challenging tasks. As illustrated in Fig.~\ref{fig_19}, three potential directions can be explored for further enhancing the inference capabilities of SIM:
\begin{itemize}
\item[$\diamond$] \textsl{\textbf{DNN:}} Developing materials and meta-atom designs that exhibit nonlinear responses would be critical for creating authentic DNNs. The inference accuracy of SIMs could benefit from directly integrating a nonlinear module into the meta-atoms. For instance, by altering the voltages applied to the varactor, a programmable activation function can be obtained by detecting the input intensity and by feeding back the threshold to an amplifier in each meta-atom~\cite{NE_2023_Gao_Programmable}. However, this demands appropriate modeling of the electromagnetic responses and a scalable as well as power-efficient design.
\item[$\diamond$] \textsl{\textbf{HOENN:}} The cascade of a SIM (optical network) and a shallow electronic ANN could create a hybrid optical-electronic neural network (HOENN) having improved performance~\cite{SA_2021_Li_Spectrally}. A SIM can convert phase-encoded information into intensity patterns, and then energy-efficient energy detectors can act as activation functions for enhancing the inference capability of HOENNs.
\item[$\diamond$] \textsl{\textbf{RNN:}} Existing SIM designs typically use anti-reflective coatings to suppress back-and-forth reflection losses within the SIM~\cite{LSA_2022_Luo_Metasurface}. However, taking advantage of these inter-layer reflections could enable the design of recurrent neural networks (RNNs), significantly improving the inference capability of SIMs. Characterizing the coupling between densely packed metasurface layers is quite challenging, as it depends on the configuration of other metasurface layers.
\end{itemize}

\section{Conclusions}
SIM technology holds the promise of supporting analog electromagnetic signal processing at the speed of light. In this overview, we have discussed recent advances in SIM research and development. Beginning with the fundamental principles behind SIM, we have reviewed the substantial progress made in fabricating SIMs for various inference tasks across communication, sensing, and computing applications. We have also unveiled a wealth of new opportunities that could further enhance the computing capability of SIMs and have identified key challenges that still have to be addressed. In summary, SIM has the potential to revolutionize interdisciplinary research at the intersection of electromagnetism, AI, and metamaterials science. Taking advantage of the complementary nature of these emerging technologies may significantly advance all these fields.

\nomenclature{2D}{Two-dimensional}
\nomenclature{3D}{Three-dimensional}
\nomenclature{A3C}{Asynchronous advantage actor critic}
\nomenclature{ADC}{Analog-to-Digital Converter} 
\nomenclature{ADMM}{Alternating Direction Method of Multipliers} 
\nomenclature{AI}{Artificial Intelligence} 
\nomenclature{ANN}{Artificial Neural Network} 
\nomenclature{AO}{Alternating Optimization} 
\nomenclature{AP}{Access Point} 
\nomenclature{BC}{Broadcast Channel} 
\nomenclature{BCD}{Block Coordinate Descent} 
\nomenclature{BS}{Base Station} 
\nomenclature{BER}{Bit Error Rate} 
\nomenclature{BFGS}{Broyden-Fletcher-Goldfarb-Shanno}
\nomenclature{CDMA}{Code Division Multiple Access} 
\nomenclature{CE}{Cross Entropy} 
\nomenclature{CNN}{Convolutional Neural Network} 
\nomenclature{CPU}{Central Processing Unit} 
\nomenclature{CQL}{Conservative Q-Learning}
\nomenclature{CRB}{Cramér-Rao Bound}
\nomenclature{CSI}{Channel State Information}
\nomenclature{DAC}{Digital-to-Analog Converter} 
\nomenclature{D4PG}{Distributed Distributional Deep Deterministic Policy Gradient} 
\nomenclature{DDPG}{Deep Deterministic Policy Gradient} 
\nomenclature{DFT}{Discrete Fourier Transform} 
\nomenclature{DNN}{Diffractive Neural Network} 
\nomenclature{DL}{Downlink} 
\nomenclature{DOA}{Direction-Of-Arrival} 
\nomenclature{DoF}{Degree-of-Freedom} 
\nomenclature{DPC}{Dirty Paper Coding} 
\nomenclature{DQN}{Deep Q-Network} 
\nomenclature{DRL}{Deep Reinforcement Learning} 
\nomenclature{ECO}{Electromagnetic Collaborative Object} 
\nomenclature{EE}{Energy Efficiency} 
\nomenclature{ELAA}{Extremely Large-scale Antenna Array} 
\nomenclature{FDTD}{Finite-Difference Time-Domain} 
\nomenclature{FEM}{Finite-Elements Method} 
\nomenclature{FP}{Fractional Programming} 
\nomenclature{FPGA}{Field Programmable Gate Array} 
\nomenclature{GA}{Gradient Ascent} 
\nomenclature{GAN}{Generative Adversarial Network} 
\nomenclature{GD}{Gradient Descent} 
\nomenclature{HOENN}{Hybrid Optical-Electronic Neural Network} 
\nomenclature{IM}{Index Modulation} 
\nomenclature{ISAC}{Integrated Sensing and Communication} 
\nomenclature{LC}{Liquid Crystal} 
\nomenclature{LEO}{Low-Earth Orbit} 
\nomenclature{LNA}{Low-Noise Amplifier} 
\nomenclature{LS}{Least Squares} 
\nomenclature{MAC}{Multiple Access Channel} 
\nomenclature{MARL}{Multi-Agent Reinforcement Learning} 
\nomenclature{MHACL}{Manifold-enhanced Heterogeneous multi-Agent Continual Learning} 
\nomenclature{MIMO}{Multiple-Input Multiple-Output} 
\nomenclature{MISO}{Multiple-Input Single-Output} 
\nomenclature{MMSE}{Minimum Mean Square Error} 
\nomenclature{MSE}{Mean Square Error} 
\nomenclature{NAC}{Natural Actor Critic} 
\nomenclature{NOMA}{Non-Orthogonal Multiple Access} 
\nomenclature{OAM}{Orbital Angular Momentum} 
\nomenclature{OFDM}{Orthogonal Frequency Division Multiplexing} 
\nomenclature{OMP}{Orthogonal Matching Pursuit} 
\nomenclature{PA}{Power Amplifier} 
\nomenclature{PAPR}{Peak-to-Average Power Ratio} 
\nomenclature{PGA}{Projected Gradient Ascent} 
\nomenclature{PLS}{Physical Layer Security} 
\nomenclature{QoE}{Quality of Experience} 
\nomenclature{QoS}{Quality of Service} 
\nomenclature{RF}{Radio-Frequency} 
\nomenclature{RIS}{Reconfigurable Intelligent Surface} 
\nomenclature{RNN}{Recurrent Neural Network} 
\nomenclature{RSMA}{Rate Splitting Multiple Acces} 
\nomenclature{SCA}{Successive Convex Approximation} 
\nomenclature{SDMA}{Spatial Division Multiple Access} 
\nomenclature{SDP}{Semidefinite Programming} 
\nomenclature{SDR}{Semidefinite Relaxation} 
\nomenclature{STAR}{Simultaneously Transmitting And Reflecting} 
\nomenclature{SIM}{Stacked Intelligent Metasurfaces} 
\nomenclature{SINR}{Signal-to-Interference-plus-Noise Ratio} 
\nomenclature{SWIPT}{Simultaneous Wireless Information and Power Transfer} 
\nomenclature{UAV}{Unmanned Aerial Vehicle} 
\nomenclature{UE}{User Equipment} 
\nomenclature{UL}{Uplink} 
\nomenclature{VNA}{Vector Network Analyzer} 
\nomenclature{WF}{Water Filling} 
\nomenclature{MPO}{Maximum a posteriori Policy Optimization}
\nomenclature{TD3}{Twin-Delayed Deep Deterministic}
\printnomenclature

\bibliography{IEEEabrv, ref}
\bibliographystyle{IEEEtran}
\end{document}